%% file: NC1-Main.tex
\documentclass[11pt,catalan]{article}
\usepackage{graphicx}
\usepackage{amsmath}
\pdfoutput=1
\usepackage{amsfonts}
\usepackage{amssymb}
\usepackage{bm}
\usepackage{cite}
\usepackage[mathscr]{euscript}

\usepackage[english]{babel}
\usepackage[latin1]{inputenc}
\usepackage{times}
\usepackage[T1]{fontenc}

\newcommand{\RR}{\ensuremath{ \mathbb{R}} }

\newcommand{\D}{{\rm d}}

\newcommand{\tildeM}{\tilde{M}}

\newcommand{\mU}{{\mathbf{U}}}
\newcommand{\mk}{{\mathbf{k}}}
\newcommand{\mq}{{\mathbf{q}}}
\newcommand{\Mp}{{\mathbf{p}}}
\newcommand{\ma}{{\mathbf{a}}}
\newcommand{\mx}{{\mathbf{x}}}

\newcommand{\Mu}{{\mathbf{u}}}
\newcommand{\mz}{{\mathbf{z}}}

\begin{document}
\title{Nonlocal Lagrangian fields and the second Noether theorem.\\
Non-commutative $U(1)$ gauge theory}
\author{Carlos Heredia\thanks{e-mail address: carlosherediapimienta@gmail.com} \, and \, Josep Llosa\thanks{e-mail address: pitu.llosa@ub.edu}\\
Facultat de F\'{\i}sica (FQA and ICC) \\ Universitat de Barcelona, Diagonal 645, 08028 Barcelona, Catalonia, Spain }
\maketitle

\begin{abstract}
This article focuses on three main contributions. Firstly, we provide an in-depth overview of the nonlocal Lagrangian formalism. Secondly, we introduce an extended version of the second Noether's theorem tailored for nonlocal Lagrangians. Finally, we apply both the formalism and the extended theorem to the context of non-commutative U(1) gauge theory, including its Hamiltonian and quantization, showcasing their practical utility.
\noindent
\end{abstract}
\input{NCU1-Noether}

\input{NC1-Theory}

\input{NC1-Hamiltonian}

\input{NC1-Conclusion}

\input{NC1-Appendix}

\bibliographystyle{utphys}
{\footnotesize\bibliography{References}}

\end{document}

%% file: NCU1-Noether.tex
\section{Introduction \label{S1}}
Nonlocal Lagrangian field theories are characterized by their formulation through a Lagrangian density that functionally depends on the field. This functional dependence is manifested either in the form of infinite series of derivatives or integral operators, such as the convolution one. In the scope of this article, we will focus exclusively on the nonlocality defined by integral operators. This choice stems from observations made in \cite{Heredia2023_1,Carlsson2016}, which point out that the functional domain in which infinite derivative operators are defined is more restricted compared to integral operators.

In \cite{Heredia2022, Heredia_PhD}, a novel Lagrangian and Hamiltonian formalism for nonlocal field theories is presented, representing an enhancement over previous methods like the (1+1)-dimensional Hamiltonian formalism \cite{Llosa1994,Gomis2001}. This novel nonlocal Lagrangian formalism, represented by integral operators instead of an infinite series of derivatives, yields two significant results: the development of a Hamiltonian framework for these Lagrangians and an extension of Noether's first theorem, which is particularly valuable in the derivation of the canonical and Belinfante-Rosenfeld energy-momentum tensors \cite{Heredia1, Heredia2}. 

An outline of the general ideas inspiring the generalization of the Legendre transformation to the nonlocal scenario is as follows: In the context of a standard first order Lagrangian, the role played by the momenta $\displaystyle{p_j(q,\dot{q}) = \frac{\partial L(q,\dot{q})}{\partial q^j}\,}$ is twofold. Initially, they appear as the prefactors of the variations \(\delta q^j\) within the boundary terms emerging from the application of the variational principle. Then, and as a consequence of this fact, these momenta also manifest as the prefactors of \(\delta q^j\) in the conserved quantities derived from Noether's theorem. A remarkably similar scenario occurs for a higher order Lagrangian, $L\left(q, \ldots, q^{(n)}\right)\,$, where there is a conjugate momentum $\,p_l\left(q, \ldots q^{(2n-l)}\right)\,$ for each $\, q^{(l)}\,,\;\; 0\leq l < n\,$,  contributing to the conserved current through $\,\displaystyle{\sum_{l=0}^{n-1} p_l\,\delta q^{(l)}}\,$. In the regular case,  the relations, $\,p_l\left(q, \ldots q^{(2n-l)}\right)\,$, can be inverted for the higher velocities, $q^{(l)}\,,\,\; n\leq l \leq 2n-1\,$, starting from the highest-order momentum $p_{n-1}\left(q, \ldots q^{(n)}\right)\,$ and so on. 

In the context of Lagrangians that depend on derivatives of infinite order, the canonical momenta are typically formulated by substituting $n$ with $\infty$ and transforming finite sums into infinite series. However, the absence of a highest-order derivative prevents the straightforward inversion of the generalized Legendre transformation --- specifically, the substitution of one half of the derivatives with the canonical momenta becomes infeasible.  To put it bluntly, because ``half of infinity is infinity". Consequently, the Legendre transformation inevitably becomes singular in the context of infinite order. Moreover, the Euler-Lagrange equations form a differential system of infinite order, for which there is no established theorem guaranteeing existence and uniqueness; this leaves the issue of initial conditions highly ambiguous, specifically regarding the quantity of initial data required to uniquely determine a solution.

Given these complexities, a paradigm shift is required: the dynamic space cannot be the space of initial data, as it is not even clear that such a concept is applicable or meaningful in this context. Instead, our attention shifts towards the space of dynamic trajectories, constituting a subspace of $\mathcal{K}$, the class of the kinematic trajectories. This specific subspace is based on those trajectories that satisfy the Euler-Lagrange equations. As a consequence of this fact, it establishes these equations as constraints that define the dynamic space. Simultaneously, canonical momenta also play the role of constraints that, together with the Euler-Lagrange equations, define the dynamic space $\mathcal{D}$ as a submanifold --- maybe infinite dimensional --- within the cotangent space $T^\ast\mathcal{K}\,$.

The standard canonical structure on $T^\ast\mathcal{K}\,$, characterized by the canonical coordinates $q^{(l)}$ and $p_k\,$, $\;l,k=0,1 \ldots\,,$ can be translated into a canonical formalism on $\mathcal{D}\hookrightarrow T^\ast\mathcal{K}\,$, the submanifold defined by the Euler-Lagrange equations and their derivatives, along the momenta, $\,p_l = p_l\left(q, \ldots q^{(k)} \ldots\right)\,$, taken as constraints. This can be achieved via two methodologies: by employing the Dirac brackets for the aforementioned constraints  \cite{Gomis2001}, or equivalently, by pulling the symplectic form $\,\displaystyle{\Omega =\sum_{l=0}^\infty \D p_l \wedge \D q^{(l)}}\,$ back onto $\mathcal{D}\,$ \cite{Heredia2}. A quite similar scenario occurs with Lagrangians that are functionally dependent on the entire trajectory, $\,q(\tau),\;\, \tau\in \RR$ (or a finite segment of it), where the canonical momentum\footnote{We use the singular but it corresponds to a continuous infinity of values, one for each $\tau$.}, $P([q],\tau)\,,$ is a functional that depends on the whole $q$, i.e. nonlocal. Once more, the Euler-Lagrange equations and the canonical momentum $p(\sigma)= P([q],\sigma)$ must be taken as constraints to define the dynamic space $\mathcal{D}$ as a submanifold on $\,T^\ast\mathcal{K}\,$, where $\mathcal{K}\,$ represents the space of all possible trajectories. The technicalities of this procedure are developed in detail in \cite{Heredia2022, Heredia_PhD}.

Most infinite-order Lagrangians in literature are obtained from Lagrangians that depend functionally on the whole trajectory. This transformation involves substituting $q(\tau)$ by the formal Taylor series 
$$ q(\tau+t) = \sum_{n=0}^\infty \frac{\tau^n}{n!}\,q^{(n)}(t) \,. $$ 
As kinematical trajectories are not necessarily real analytic, this relationship possesses only heuristic significance. Nevertheless, it serves as a bridge connecting the treatment involving infinite derivatives with the functional approach. Hence, in light of this consideration, throughout this manuscript, we will consistently employ this novel formalism based on integral operators, rather than the infinite series approach. 

In \cite{Tomboulis2015} Tomboulis sets up an alternative Hamiltonian framework for nonlocal Lagrangians characterized by integral operators. This framework is expressly limited to kernels with compact support, referred to as quasilocal. This is a noteworthy distinction between his proposal and our current approach, based on \cite{Heredia2022, Heredia_PhD}, which does not suffer from this limitation. Particularly, the case study we consider here, from Section 4 onwards, is fully nonlocal.

A relevant illustration of nonlocality can be found in the framework of non-commutative (NC) spacetime \cite{JaumeGomis2000}, particularly in the context of U(1) gauge theory \cite{Gomis2001_2, Naser2015, Kupriyanov2020, Kupriyanov2021}. The fundamental motivation behind introducing non-commutativity into this theory (apart from String Theory's motivation) is to enhance renormalizability at short distances and potentially establish finiteness within these scales \cite{Douglas_NCFT}. See, for instance, \cite{Kupriyanov2020_2, Marija2023, Trampetic2023} (and references therein) for other non-commutative theories. 

The paper is structured into two main parts. In the first part ---Sections \ref{S2} to \ref{S3}---, we extend the second Noether theorem to nonlocal Lagrangians and apply it to NCU(1) gauge theory for testing purposes. In the second part ---from Section \ref{S3.5} to the end---, we practically implement the Hamiltonian formalism developed in \cite{Heredia2022} within the framework of NCU(1) gauge theory.

The article is organized in seven sections. Section \ref{S2} is an outline of the nonlocal Lagrangian formalism already developed in previous works \cite{Heredia2,Heredia2022,Heredia_PhD}: we present the variational principle for such a Lagrangian and derive the nonlocal Euler-Lagrange field equations\footnote{In the present work we focus on nonlocal Lagrangians that do not depend explicitly on the spacetime point because this is the case of the NCU(1) gauge theory to which we intend to apply our mathematical methods.}. 
We also pay some attention to these field equations when the nonlocal Lagrangian is a total divergence. As proved in \cite{Heredia_PhD} and contrarily to what happens in the local case, these field equations do not identically vanish. We here outline a sufficient condition \cite{Heredia2022} for the nonlocal Euler-Lagrange field equations derived from a nonlocal total divergence do identically vanish. 
In Section \ref{Noe}, we show Noether's first theorem and extend Noether's second theorem to encompass nonlocal gauge symmetries, being local symmetries a specific case of this broader framework. Next, in Section \ref{S3}, we focus on the NCU(1) gauge theory; we explore how the nonlocal Lagrangian and the field equations transform under a nonlocal gauge transformation, and analyse the extended Noether's second theorem in the context of this theory. The nonlocality inherent in this theory, i. e. the Moyal $*$ bracket, is addressed through an integral operator rather than the previously used infinite series \cite{Seiberg1999, Gomis2001_2}.

In Section \ref{S3.1a},  we deepen the study of the dynamical space in the context of the NCU(1) gauge theory and propose a solution for the field equations using a perturbative approach. Furthermore, we somehow revisit the Seiberg-Witten map within this specific context. In Section \ref{S3.5}, we implement the Hamiltonian formalism introduced in \cite{Heredia2022} into the NCU(1) gauge theory. This leads us to derive the symplectic form, identify the canonical variables, establish the elementary Poisson brackets, determine the Hamiltonian up to the second order in some nonlocality parameter, and adress the canonical quantization. Additionally, for readers who are more interested in reproducing the calculations, we provide detailed steps in Appendices A and B to guide them through the process.

\section{An outline on nonlocal Lagrangian field theories \label{S2}}
In this section we review the nonlocal Lagrangian formalism proposed in \cite{Heredia2022,Heredia2} to which the interested reader may be referred for further details. To begin with, we introduce the notions of kinematic space, dynamic space, nonlocal Lagrangian density and spacetime translation. 

The kinematic space is the class of all possible fields $\mathcal{C}^m\left(\RR^4, \RR^n\right)\,$ that may not necessarily be on shell, whereas the dynamic space $\mathcal{D}\,$ is the subclass of fields fulfilling the field equations\footnote{For the sake of conciseness, we consider a four-dimensional spacetime $\mathbb{R}^{3+1}$ but the results here derived are also valid for any number of dimensions.}. 

A nonlocal Lagrangian density is a real-valued functional on the kinematic space $\mathcal{K}\,$: 
$$ (\phi^A) \in \mathcal{K} \longrightarrow \mathcal{L}(\phi^A) \in \RR \,,$$
which may depend on all values $\phi^A(y)$, $\,A=1 \ldots m,\,$ of the field variables at any point $y^b$. This is why it is called nonlocal. Notice that the functional dependence is not emphasized with square brackets, as is common elsewhere. We will also dispense with superscripts for both field variables and point coordinates, unless the context explicitly requires it.

Given $y\in\RR^4$, the spacetime translation is the transformation
\begin{equation}  \label{L1} 
  (\phi) \stackrel{T_y}{\longrightarrow} (T_y \phi) \,,\qquad {\rm where} \qquad T_y \phi(z) = \phi(z+y)\,.
\end{equation}
The class of all these transformations is an Abelian group 
$$\; T_{ y_1}\circ T_{ y_2} = T_{ y_1+ y_2}\,.$$ 
Notice that all the information about the field evolution is already contained in $\phi(z)\,.$

As presented in ref. \cite{Heredia2022}, the variational principle is based on the one-parameter family of finite action integrals 
\begin{equation} \label{A1}
  S(\phi,R) = \int_{|y|\leq R} \D y\,\mathcal{L}(T_y\phi)  \,, \qquad \forall R\in \RR^+ \,,
\end{equation}
where $\,|y|= \sqrt{\sum_{j=1}^4 (y^j)^2 }\;$ is the Euclidean length, and it reads
\begin{equation} \label{A2}
\lim_{R\rightarrow\infty} \delta S(\phi,R) \equiv \lim_{R\rightarrow\infty}  \int_{|y|<R} \D y\,\int_{\RR^4} \D z\,\frac{\delta \mathcal{L}\left(T_y \phi\right)}{\delta \phi(z)} \,{\delta \phi(z)} = 0\,,  
\end{equation} 
for all variations $\,\delta \phi(z) \,$ with compact support. The Lagrange field equations are 
\begin{equation}  \label{L2o} 
\psi_A(\phi,z) = 0 \,, \quad {\rm with} \quad \psi_A(\phi,z) := \int_{\RR^4} \D  y\,\lambda_A(\phi, y,z) \quad {\rm and} \quad
\lambda_A(\phi, y,z) := \frac{\delta \mathcal{L}\left(T_y\phi\right)}{\delta \phi^A(z)} \,,
\end{equation}
and the dynamic fields are the solutions of these equations. 
From these definitions it easily follows that
\begin{equation}  \label{L2oz} 
\psi_A(\phi,x+z) = \psi_A(T_x\phi,z)  \qquad \quad {\rm and} \qquad \quad
\lambda_A(\phi, x+y, x+z) = \lambda_A(T_x\phi, y, z) \,.
\end{equation}

\subsection{The Lagrange equations for a total divergence \label{S2.1a}}
A common feature of local theories is that, if the Lagrangian density is a total divergence,
\begin{equation}  \label{TD1}
 \mathcal{L}(x) = \partial_b W^b(x) \,,
\end{equation}
then the Lagrange equations vanish identically. In our notation, $\mathcal{L}(x)$  and $W^b(x)$ must be respectively understood as $\mathcal{L}(T_x\phi)$ and $W^b(T_x\phi, x)$.

In the nonlocal context things are not so simple because, once the requirement of locallity is relaxed, equation (\ref{TD1}) has infinitely many solutions. Indeed, its general solution is: 
$$ W^b(x) = \delta_4^b\,\int_\RR \D \tau\,\left[\theta(\tau)-\theta(\tau-x^4)\right]\,\mathcal{L}(\mathbf{x},\tau) + \partial_c\Omega^{bc}(x) \,,$$ 
where $\,\mathbf{x}=(x^1,x^2,x^3)\,$ and $\Omega^{bc}+\Omega^{cb} = 0\,$. 
Nevertheless this does not imply that the Lagrange equations for any $\mathcal{L}$ are identically null, as it is implied in the context of standard local Lagrangians and local $W^b\,$ ---see an example in \cite{Heredia_PhD}.

The family of actions (\ref{A1}) for such a Lagrangian $ \partial_b W^b(\phi,x) $ is
$$   S(\phi,R) = \int_{|y|<R} \D y\,\partial_b W^b(T_y\phi,y)  = \int_{|y|=R} \D \Sigma_b(y)\,W^b (T_y \phi,y)\,,$$
where Gauss' theorem has been applied and $\,\D \Sigma_b(y)\,$ is the hypersurface element on $\,|y|=R\,$. Thus, the variational principle (\ref{A2}) yields the field equations
$$ \psi(\phi,z) :=\lim_{R\rightarrow\infty} \frac{\delta S(\phi,R)}{\delta \phi(z)} \equiv 
\lim_{R\rightarrow\infty} \int_{|y|=R} \D \Sigma_b(y)\,\frac{\delta W^b (T_y \phi,y)}{\delta\phi(z)}\,,$$ 
which do not necessarily vanish.
However, as $\, \D \Sigma_b(y)\,$ scales as $|y|^3$, they do vanish identically provided that 
\begin{equation} \label{TD3}
 \lim_{|y|\rightarrow\infty} \left\{|y|^3\,\frac{\delta W^b (T_y \phi,y)}{\delta \phi(z)} \right\} \equiv 0  \,,
\end{equation}
which yield a sufficient condition for the Lagrangian $\partial_b W^b(x)$ to yield identically null Lagrange equations. This condition is obviously met if $ W^b (T_y \phi,y)$ is local.

\section{Noether's theorem  \label{Noe}}
Consider the infinitesimal transformation
\begin{equation}  \label{A3}
x^{\prime a}(x) = x^a + \delta x^a(x) \,, \qquad \quad \phi^{\prime A}(x) = \phi^A(x) + \delta \phi^A(x) \,.
\end{equation}
The transformation of the Lagrangian density is defined so that the action integral over any four-volume is preserved, namely,
\begin{equation}  \label{N1}
 S(\mathcal{V})=S^\prime (\mathcal{V}^\prime)\,, \quad {\rm with} \quad  S(\mathcal{V}) \equiv \int_{\mathcal{V}} \D x \,\mathcal{L}\left(T_x\phi^A \right)  \quad {\rm and} \quad  
S^\prime (\mathcal{V}^\prime) \equiv \int_{\mathcal{V}^\prime} \D  x^\prime\, \mathcal{L}^\prime\left(T_{ x^\prime}\phi^{\prime A} \right) \,,
\end{equation}
where ${\mathcal{V}^\prime}$ is the transformed of the spacetime volume $\mathcal{V}\,$.
Therefore, $\,\displaystyle{ \mathcal{L}^\prime\left(T_{ x^\prime}\phi^{\prime A} \right) = \mathcal{L}\left(T_x\phi^A \right) \,\left|\frac{\partial x}{\partial x^\prime}\right| \,,}$
and we define
\begin{equation}  \label{N2}
  \delta \mathcal{L}\left(\phi^A \right) :=  \mathcal{L}^\prime\left(\phi^A \right) - \mathcal{L}\left(\phi^{A} \right) \,.
\end{equation}

From (\ref{N1}) it follows that 
$\,\displaystyle{\int_{\mathcal{V}^\prime} \D  x\,\mathcal{L}^\prime\left(T_x\phi^{\prime A}\right) -\int_{\mathcal{V}}\D  x\,\mathcal{L}\left(T_x\phi^A \right)  \equiv 0 }\,$,
where we have replaced the dummy variable $ x^\prime$ with $ x\,$. Neglecting second order infinitesimals, we get that
\begin{equation}  \label{A5}
 \int_{\mathcal{V}} \D x\,\left[ \mathcal{L}^\prime\left(T_x\phi^{\prime A}\right) - \mathcal{L}\left(T_x\phi^A \right)\right]\, + \int_{\partial \mathcal{V}} \mathcal{L}(T_x\phi^A )\,\delta  x^b \,\D \Sigma_b = 0\,.
\end{equation}
Then, including (\ref{N2}), we can write
\begin{eqnarray*}
 \mathcal{L}^\prime\left(T_x\phi^{\prime A} \right) - \mathcal{L}\left(T_x\phi^A \right) & = &\mathcal{L}^\prime\left(T_x\phi^{\prime A} \right) - \mathcal{L}\left(T_x\phi^{\prime A} \right) + \mathcal{L}\left(T_x\phi^{\prime A} \right) - 
\mathcal{L}\left(T_x\phi^A \right) \\[1ex]
  & = & \delta \mathcal{L}\left(T_x\phi^A \right) + \int_{\RR^4} \D y\,\frac{\delta \mathcal{L}\left(T_x\phi \right) }{\delta \phi^A(y)}\,\delta\phi^A(y) + O(2)\,,
\end{eqnarray*}
where $\delta\phi^A(y) := \phi^{\prime A}(y) - \phi^A(y)\,$ and we have used that, up to second order infinitesimals, $\,\delta \mathcal{L}(T_x\phi^{\prime A}) = \delta \mathcal{L}(T_x\phi^A)\,$. Therefore,
\begin{equation}  \label{A5z}
  \mathcal{L}^\prime\left(T_x\phi^{\prime A} \right) - \mathcal{L}\left(T_x\phi^A \right) = \delta \mathcal{L}\left(T_{ x}\phi^A \right) + \int_{\mathbb{R}^4} \D  y\,\lambda_A(\phi, x, y)\,\delta \phi^A( y)  \,, 
\end{equation}
where $\,\lambda_A(\phi, x, y)\,$ is defined in (\ref{L2o}). By introducing the variable $z = y - x$ in the latter equation and then substituting it into (\ref{A5}) while applying Gauss' theorem, we arrive at
$$  \int_{\mathcal{V}}\D  x\, \left\{ \partial_b\left[\mathcal{L}(T_x\phi^A )\,\delta x^b \right] + \delta \mathcal{L}\left(T_x\phi^A \right) + \int_{\RR^4} \D z\,\lambda_A(\phi, x,z+ x)\,\delta\phi^A(z+ x) \right\} = 0 \, $$
that, including (\ref{L2o}), can be rewritten as
\begin{eqnarray} 
\lefteqn{ - \int_{\mathcal{V}}\D  x\,\psi_A(\phi, x) \,\delta\phi^A( x) = \int_{\mathcal{V}}\D  x\, \left\{\partial_b\left(\mathcal{L}(T_x\phi^A)\,\delta  x^b\right) + \delta \mathcal{L}\left(T_x\phi^A\right)     \right.} \nonumber \\[2ex]   \label{A6}
 & & \left. \hspace*{4em}+\int_{\mathbb{R}^4} \D z \,\left[ \lambda_A(\phi, x,z+ x)\,\delta\phi^A(z+ x) - \lambda_A(\phi, x-z, x)\,\delta\phi^A( x)   \right]\right\}\,.
\end{eqnarray}

Using now the identity
\begin{eqnarray*}
\lefteqn{\lambda_A(\phi, x,z+ x)\,\delta\phi^A(z+ x) - \lambda_A(\phi, x-z, x)\,\delta\phi^A( x) =} \\[1ex]
 &  =& \int_0^1 \D s \,\frac{\D\;}{\D s}\left\{\lambda_A(\phi, x+[s-1] z, x+s z)\,\delta\phi^A( x+s z) \right\}   \\[1ex]
 & =& \int_0^1 \D s\,z^b\,\frac{\partial\quad}{\partial x^b}\left\{\lambda_A(\phi, x+[s-1]z, x+s z)\,\delta\phi^A( x+s z)\right]\} \\[1ex]
 & =& \int_0^1 \D s \, z^b \,\frac{\partial\quad}{\partial x^b} \left\{\lambda_A(T_x\phi, [s-1]z, s z)\,\delta T_x\phi^A( s z) \right]\} 
\end{eqnarray*}
in equation  (\ref{A6}), we obtain
\begin{equation}  \label{A7}
  \int_{\mathcal{V}}\D  x\,\left\{\psi_A(\phi, x) \,\delta\phi^A( x) + \delta\mathcal{L} + \frac{\partial\quad}{\partial  x^b} \left[\mathcal{L}\,\delta x^b + \Pi^b(T_x\phi) \right] \right\} = 0\,,
\end{equation}
where $\,\mathcal{L}\,$ and $\,\delta\mathcal{L}\,$ are shorthands for $\,\mathcal{L}(T_x\phi)\,$ and $\,\delta\mathcal{L}(T_x\phi)\,$, and 
\begin{equation}  \label{A8}
\Pi^b(\phi) :=  \int_{\RR^4} \D z\, z^b \int_0^1 \D s\,\lambda_A(\phi, [s-1] z, s z)\,\delta\phi^A(s z)\,,  
\end{equation}
where (\ref{L2oz}) has been included.

Now, as equation (\ref{A7}) holds  for any spacetime volume $\mathcal{V}$, it follows that
\begin{equation}  \label{A9}
\frac{\partial\quad}{\partial  x^b} \left[J^b(T_x\phi, x) \right] + \delta\mathcal{L}(T_x\phi)  + \psi_A(\phi, x) \,\delta\phi^A( x) \equiv 0  \,,
\end{equation}
where
\begin{equation}  \label{A10}
\quad J^b(T_x\phi,x) := \mathcal{L}(T_x\phi)\,\delta x^b(x) + 
\int_{\RR^4} \D z\,z^b \int_0^1 \D s\,\lambda_A(T_x\phi,[s-1] z, s z)\,\delta T_x \phi^A(s z)\,. 
\end{equation}
The identity (\ref{A9}) holds for any Lagrangian and any infinitesimal transformation, regardless of whether the field equations are invariant or not.  

The transformation (\ref{A3}) is said a {\em nonlocal Noether symmetry} if
\begin{equation}  \label{A4a}
  \delta \mathcal{L}\left(T_{ x}\phi^A \right) = \partial_b W^b\left(T_x\phi^A, x\right) \,,
\end{equation}
that is $\delta \mathcal{L}$ is the four-divergence for some $\,W^b\,$ fulfilling the asymptotic condition (\ref{TD3}). 
Hence for a Noether symmetry equation (\ref{A9}) becomes
\begin{equation}  \label{A9a}
\frac{\partial\quad}{\partial x^b} \mathcal{J}^b(T_x\phi,x) +\psi_A(\phi, x)\,\delta\phi^A( x) \equiv 0 \,,
\qquad {\rm with} \qquad \mathcal{J}^b(T_x\phi,x) := J^b(T_x\phi,x) + W^b(T_x \phi,x)\,,
\end{equation}
which is known as the  Noether identity and it holds for any kinematic field $\phi^A\,$. Only for dynamic fields, i. e. {\em on-shell}, this identity implies the local conservation of the current:
\begin{equation}  \label{A10a}
  \partial_b \mathcal{J}^b(T_x\phi,x) = 0 \,.
\end{equation}

\subsection{Finite dimensional Lie groups of transformations. First Noether theorem}
Let the transformations (\ref{A3}) belong to an $N$-parameter Lie group of Noether symmetries, 
$$ \delta x^b(x) = \varepsilon^\alpha \,\xi_\alpha^b(x) \,,\qquad \delta\phi^A(x) = \varepsilon^\alpha\,\Phi_\alpha^A(\phi,x) \,,\qquad W^b(T_x\phi,x) = \varepsilon^\alpha \,W_\alpha^b(T_x\phi,x) \,, $$
where $\,\xi_\alpha^b(x)\,$ is the infinitesimal generator for the constant parameter $\varepsilon^\alpha\,,\,\,\alpha = 1 \ldots N\,$ and 
$\,\delta\phi^A\,$ might depend functionally on $\,\phi^B\,$. Then, the conserved current $\mathcal{J}^b(T_x\phi,x)$ can be written as
$$ \mathcal{J}^b(T_x\phi,x) = \varepsilon^\alpha \mathcal{J}_\alpha^b(T_x\phi,x) \,,\qquad {\rm with} \qquad \partial_b \mathcal{J}_\alpha^b(T_x\phi,x) = 0 \,. $$
Hence, for each group parameter there is a conserved current given by
\begin{equation}  \label{A10b}
 \quad \mathcal{J}_\alpha^b(T_x\phi,x) := \mathcal{L}\,\xi_\alpha^b(x) + W_\alpha^b +  \int_{\RR^4} \D z\,z^b \int_0^1 \D s\,\lambda_A(T_x\phi,[s-1] z, s z)\,\Phi_\alpha^A(T_x\phi,s z)  \,.
\end{equation}
Recall that $\,\mathcal{L}\,$ and $\,W_\alpha^b\,$ respectively refer to $\,\mathcal{L}(T_x\phi)\,$ and $\,W_\alpha^b(T_x\phi, x)\,$.

\subsection{Hamiltonian formalism \label{SS3.2}} 
 
Assuming the Lagrangian remains invariant under the variation $\delta \phi^A\,$, i. e. $W^b = 0\,$, the local conservation law (\ref{A10a}) reads $\,\partial_b J^b = 0\,$ and, provided that the fields $\,\phi^A(\mx, t)\,$ decay fast enough at spatial infinity, the total charge
\begin{equation} \label{AM0}
Q(t) := \int_{\RR^3} \D\mx\,J^4(T_x\phi, \mx,t) 
\end{equation}
is a constant. Therefore, using (\ref{A10}), we get that
\begin{equation*} 
Q(t) = \int_{\RR^3} \D\mx\,\mathcal{L}\left(T_x\phi\right)\,\delta x^4(\mx,t) +
\int_{\RR^6} \D\mx\,\D\mz\,\int_\RR \D \zeta\,\zeta \int_0^1 \D s \,\lambda_A(\phi,x+(s-1) z, x+ sz)\,\delta \phi^A(x+sz)\,.
\end{equation*} 
As highlighted in the Introduction, our focus in defining the canonical momentum centers specifically on the prefactor of $\delta \phi^A$, drawing an analogy with the local case. By introducing a set of new variables $\mathbf{z}$, $\Mu = \mathbf{x} + s \mathbf{z}$, $\zeta$, and $\rho = s \zeta$, we can re-express the previously equation in terms of these variables as follows:
\begin{equation}  \label{AM1}
Q(t) = \int_{\RR^3} \D\mx\,\mathcal{L}\left(T_x\phi\right)\,\delta x^4(\mx,t) +
\int_{\RR^6} \D\Mu\,\D\mz\,\int_\RR \D \zeta  \int_0^\zeta \D \rho \,\lambda_A(\phi,\Mu- \mz,t+\rho-\zeta, \Mu,\rho)\,\delta \phi^A(\Mu,\rho)
\end{equation} 
and, using that
$$ \int_\RR \D \zeta \int_0^\zeta \D \rho \; \ldots = \int_{\RR^2} \D\rho\,\D\zeta \,\left[\theta(\rho)- \theta(\rho-\zeta)\right]\;\ldots \,,$$
we arrive at
\begin{equation*}  
Q(t) = \int_{\RR^3} \hspace*{-.75em} \D\mx\,\mathcal{L}\left(T_x\phi\right)\,\delta x^4(\mx,t) +
\int_{\RR^4}  \hspace*{-.75em} \D u\,\delta \phi^A(\Mu,\rho)\,\int_{\RR^4}  \hspace*{-.75em} \D z\,\left[\theta(u^4)- \theta(u^4-z^4)\right]\,\lambda_A(\phi,u^b- z^b +t \delta_4^b,u)\,,
\end{equation*}
where $u^b = (\Mu,\rho)\,$ and $z^b = (\mz,\rho)\,$. 

Finally, since the charge is conserved: $Q(t)= Q(0)\,,$ we get that
\begin{equation}  \label{AM2}
Q = \int_{\RR^3} \D\mx\,\mathcal{L}\left(T_\mathbf{x}\phi\right)\,\delta x^4(\mx,0) + \int_{\RR^4} \D u\,P_A(\phi, u)\,\delta \phi^A(u) \,,
\end{equation} 
with
\begin{equation}  \label{AM3}
P_A(\phi,u) = \int_{\RR^4} \D z\,\left[\theta(u^4)- \theta(u^4-z^4)\right]\,\lambda_A(\phi,u - z, u)\,,
\end{equation}
which we shall adopt as the definition of the canonical momentum.

The Hamiltonian, representing the generating function for time evolution, corresponds to the conserved charge derived from the variations $\delta x^b = - \varepsilon \delta_4^b\, $ and $\,\delta \phi^A =\varepsilon\,\partial_4\phi^A\,$, namely, the total energy. The energy density, identified as the time component of the current (\ref{A10b}), is:
$$ \mathcal{E} = -\mathcal{L}(T_x\phi) + 
\int_{\RR^3}\D\mz\,\int_\RR \D \zeta\,\zeta \int_0^1 \D s \,\lambda_A(\phi,x+(s-1) z, x+ sz)\,\delta \phi^A(x+sz)\,. $$
By following the same steps described above, or by directly applying the variations mentioned earlier to equation (\ref{AM2}), we obtain that the Hamiltonian (total energy) $h:=Q/\epsilon$ is:
\begin{equation}  \label{AM4}
h =  \int_{\RR^4} \D u\,P_A(\phi, u)\,\partial_4\phi^A(u) - L(\phi) \,,
\end{equation} 
where 
$$ L(\phi):=  \int_{\RR^3} \D \mathbf{x}\, \mathcal{L}(T_{\mx}\phi)\,.  $$ 
As proven in \cite{Heredia2022, Heredia_PhD}, the Hamilton equations (in geometric form) are $\,i_\mathbf{H} \omega = - \delta h\,$, where 
\begin{equation}\label{eq:SF}
 \omega = \int_{\RR^4} \D u \, \delta P_A(\phi, u) \wedge \delta \phi^A(u)    
\end{equation}
 is the (pre)symplectic form on the dynamic space $\mathcal{D}$,  
$$ \mathbf{H}f(\phi,u) := \partial_\epsilon\left[f(T_\epsilon\phi,u)\right]_{\epsilon=0} $$
is the generator of time evolution in  $\mathcal{D}$.

\subsection{Gauge groups. Second Noether theorem}
When dealing with a gauge group, the transformations (\ref{A3}) depend on a number of arbitrary functions $\varepsilon^\alpha(y)\,$, $\,\alpha = 1 \ldots N\,$, maybe in a nonlocal manner,
\begin{equation}  \label{GG1}
\delta\phi^A (x) = \int_{\mathbb{R}^4} \D y\, R^A_\alpha(x,y)\,\varepsilon^\alpha(y) \,.
\end{equation} 
For the sake of simplicity, we will focus only on the case $\,\delta x^b = 0\,$; thus, we find that  
\begin{equation}  \label{GG1a}
 \psi_A(x) \,\delta\phi^A(x) =  \int_{\RR^4} \D z \, \psi_A(x)\, R^A_\alpha(x,x+z )\,\varepsilon^\alpha(x+z) \,,
\end{equation} 
where we have taken $\; y=x+z \,$ and, to avoid an overelaborated notation, we have written $\psi_A(x)$ and $R^A_\alpha(x,y) $ instead of $\psi_A(\phi,x)$ and $R^A_\alpha(\phi,x,y)\,$.
Using the identity
\begin{eqnarray*}
 \lefteqn{\psi_A(x)\, R^A_\alpha(x,x+z )\,\varepsilon^\alpha(x+z) - \psi_A(x-z )\, R^A_\alpha(x-z ,x)\,\varepsilon^\alpha(x) =} \\[1ex]  & & \hspace*{5em} \int_0^1\D s \,z^b \partial_b\left[ \psi_A(x+[ s -1]z )\, R^A_\alpha(x+[ s -1]z ,x+ s  z )\,\varepsilon^\alpha(x+ s  z )  \right] \,,
\end{eqnarray*}
equation (\ref{GG1a}) becomes
\begin{equation}  \label{GG2a}
\psi_A(x) \,\delta\phi^A(x) = \varepsilon^\alpha(x)\,N_\alpha(x) + \frac{\partial K^b(x)}{\partial x^b}  \,, 
\end{equation}
with 
\begin{equation}  \label{GG2b}
N_\alpha(x):= \int_{\RR^4} \D y\,\psi_A(y)\, R^A_\alpha(y,x)
\end{equation}
and
$$ K^b(x) := \int_{\RR^4} \D  z \,\int_0^1\D s \, z^b\, \psi_A(x+[s -1] z )\, R^A_\alpha(x+[ s -1] z ,x+ s z)\,\varepsilon^\alpha(x+ s z)\,. $$ 
Again, for the sake of a simple notation, we have not explicited the functional dependence of $N_\alpha$ and $K^b$ on $\phi\,$.

Upon introducing the variable $\,y=x+sz\,$, the last expression becomes
\begin{equation}  \label{GG3}
K^b(x) := \int_{\RR^4} \D y \,\varepsilon^\alpha(y)\,
\int_0^1 \frac{\D s}{s^{n+1}}\, (y^b - x^b) \, \psi_A(\xi )\, R^A_\alpha(\xi,y)\,,
\end{equation}
where $\; \displaystyle{\xi = \frac{s-1}s\,y +\frac1s\,x} \,$ and $n$ is the number of dimensions of the space of $x^b$. Finally, substituting (\ref{GG2a}-\ref{GG3}) into (\ref{A9}), we arrive at the Noether identity
\begin{equation}  \label{GG3a}
 \varepsilon^\alpha(x)\, N_\alpha(x) + \frac{\partial \left(K^b(x) + J^b(x)\right)}{\partial x^b} + \delta \mathcal{L} (x)=0  \,,
\end{equation}
where $J^b(x)$ is a shorthand for $\,J^b(T_x\phi,x)\,$. 

We have used the Noether identity in the form (\ref{A9}) rather than (\ref{A9a}), i. e. we have not replaced $\delta \mathcal{L}(x)$ with 
 $\partial_b W^b(x)$ because we are interested in covering a wider class of transformations than just Noether symmetries. The motivation of this will be seen in Section \ref{S3.2}.

Furthermore, by using (\ref{GG1}) and including that $\delta x^b = 0\,$, the current (\ref{A10}) can be written as:
\begin{equation}  \label{GG3z}
J^b(x) := \int_{\RR^4} \D y\,\varepsilon^\alpha(y)\,
\int_0^1 \frac{\D s}{s^{n+1}}\,\int_{\RR^4}\D u\, (u^b - x^b) \, \lambda_A(\tilde\xi,u )\, R^A_\alpha(u,y)\,,
\end{equation}
where the variable $\,u=x+s z\,$ has been used and $\;\displaystyle{\tilde\xi = \frac{s-1}s\,u +\frac1s\,x}  \,$.

The Noether identity (\ref{GG3a}) applies for all kinematic fields and the terms $N_\alpha$ and $K^b$ exhibit a linear dependence on $\,\psi_A(\xi)\,$. Therefore, this identity serves as a functional constraint on the field equations. 

%% file: NC1-Theory.tex
\section{The non-commutative U(1) gauge theory  \label{S3}}
Now the concepts developed in previous sections will be applied to the non-commutative $U(1)$ gauge theory. Please notice our outsider perspective: we focus on displaying the application of our mathematical methods rather than analyzing the results, a task we leave to the  experts on this field. 

Throughout this section we will refer to the results derived in Appendix A, where we have extensively detailed the transformation of the Moyal product and its bracket into integral forms, as well as the expression of the Moyal bracket in Fourier space. In addition, we provide a list of eleven properties to which we will refer the reader interested in reproducing in detail the developments in the present section.

The dynamical variable is a 1-covariant real field $A_b(x)$, $\,b= 1,\ldots, 4\,,\,x\in \RR^4$. For simplicity, we will focus on the four-dimensional scenario, but extending it to $n$ dimensions is straightforward. The nonlocal action integral is
\begin{equation}  \label{NC0}
 S(R) = \int_{|x| \leq R} \D x\,\left(-\frac14\,F^{ab}(x)\,F_{ab}(x) \right)\,,
\end{equation}
where\footnote{Indices are raised and lowered with the Minkowski metric $\;\eta_{ab} = {\rm diag}(1,1,1,-1)\,.$} 
\begin{equation}  \label{NC1}
F_{ab} := \partial_a A_b - \partial_b A_a - i [A_a, A_b]\,,
\end{equation}
and the commutator is the one associated with the Moyal $\ast$ product
\begin{equation}  \label{NC2}
[f,g] = f\ast g - g\ast f\,,\qquad (f\ast g )_{(x)} := \left[\exp\left(\frac{i}2\,\theta^{ab}\,\frac{\partial \;}{\partial \alpha^a}\,\frac{\partial \;}{\partial \beta^b}\right) \,f(x+\alpha)\,g(x+\beta)\right]_{\alpha=\beta=0} \,,
\end{equation}
where $\theta^{ab}$ is a constant, skewsymmetric, contravariant 2-tensor. In Appendix A, we show that if $\theta^{ab}$ is non-degenerate\footnote{For simplicity, we will restrict to the non-degenerate case but similar results hold for degenerate $\theta^{ab}$.}, then
\begin{equation}  \label{NC3}
(f\ast g )_{(x)} = \frac1{\pi^4|\theta|}\,\int_{\RR^8} \D u\,\D v\,f(x-u)\,g(x-v)\,e^{2 i u \tilde\theta v}
\end{equation}
and 
\begin{equation}  \label{NC4}
[f,g]_{(x)} = \int_{\RR^8} \D u\,\D v\,f(x+u)\,g(x+v)\, \tildeM(u,v)  \,, \quad {\rm where} \quad \tildeM(u,v) := \frac{2 i}{\pi^4|\theta|}\,\sin(2 v \tilde\theta u) \, ,
\end{equation}  
$v \tilde\theta u:= v^a \tilde\theta_{ab} u^b\,$, $\,\tilde\theta_{ab}\,$ is the inverse matrix for $\,\theta^{ab}\,$ and $\,|\theta|=\det(\theta^{ab})\,$. As a result, the quantities $F_{ab}(x)$ depend nonlocally on the field $\,A_b\,$. Moreover, the Moyal bracket of two real functions is an imaginary number and, as a consequence, $F_{ab}$ is real.

Introducing the covariant derivative
\begin{equation}  \label{NC6a}
D_a f := \partial_a f - i \,[A_a,f] \,,
\end{equation}
equation  (\ref{NC1}) becomes
\begin{equation}  \label{NC1a}
F_{ab} := D_a A_b - D_b A_a + i [A_a, A_b]\,,
\end{equation}
and the identity 
\begin{equation}  \label{NC1b}
 D_{c} F_{ab} + D_{a} F_{bc} + D_{b} F_{ca} \equiv 0 \,
\end{equation}
arises from the definition and properties (\ref{Ap9}) and (\ref{Ap10d}). 

The nonlocal Lagrangian density is easily inferred from the nonlocal action (\ref{NC0}) and it is  
\begin{equation}  \label{NC5}
\mathcal{L}\left(T_y A\right) = -\frac14\,F^{ab}(y)\,F_{ab}(y) \,.
\end{equation}
Next, using the definitions (\ref{NC1}) and (\ref{NC4}), and taking $\,\delta_x(y):=\delta(y-x)\,$, we obtain that 
\begin{equation}  \label{NC6}
\frac{\delta F_{ab}(y)}{\delta A_c(x)} = 2\,\delta^c_{[b} \left(D_{a]}\delta_x\right)_{(y)} \,, 
\end{equation}
which, substituted into equation (\ref{L2o}), yields
\begin{equation}  \label{NC7}
\lambda^c(y,x) = F^{ca}(y)\, \left(D_a \delta_x\right)_{(y)} \,.
\end{equation}
For simplicity, we do not make the dependence on $A_b$ explicit and use $\lambda^c(y,x)$ instead of $\lambda^c(A_b,y,x)\,$. 

The field equations arise from substituting (\ref{NC7}) into equation (\ref{L2o}) and they are
\begin{equation}  \label{NC8}
\psi^c(x) \equiv \int_{\RR^4} \D y\,F^{ca}(y)\,\left(D_a \delta_x\right)_{(y)}  \equiv D_a F^{ac}(x) = 0\,,
\end{equation}
where (\ref{NC6}) and (\ref{Ap10a}) have been included. It is worth noticing that this result coincides with that obtained in \cite{Gomis2001_2}, albeit by alternative methods.

\subsection{Gauge transformations  \label{S3.1}}
Let us consider the (infinitesimal) nonlocal gauge transformation,  
\begin{equation}  \label{NC9}
A^\prime_a(x) = A_a(x) + D_a \varepsilon(x) \,, \qquad \quad x^{\prime b} = x^b \,,
\end{equation}
which only affects the potential. The fields transform according to
\begin{equation}  \label{NC10}
F^\prime_{ab}(x) = F_{ab}(x) - i\,[F_{ab},\varepsilon]_{(x)} + O(\varepsilon^2)\,,
\end{equation}
whereas the field equations (\ref{NC8}) transform as
$$ D^\prime_a F^{\prime\,ac} \equiv D_a F^{ac} - i\,\left[D_a F^{ac}, \varepsilon\right] + O(\varepsilon^2)\,,$$ 
namely,
\begin{equation}  \label{NC11}
\psi^{\prime c} \equiv \psi^c - i [\psi^c,\varepsilon] + O(\varepsilon^2)\, \qquad {\rm or} \qquad \delta \psi^c =  - i [\psi^c,\varepsilon]\,.
\end{equation}
In turn, the variation of the Lagrangian (\ref{NC5}) is
\begin{equation}  \label{NC12}
\delta\mathcal{L}(T_x A_a) = - \frac{i}2 \,F^{ab}(x)\,[F_{ab},\varepsilon]_{(x)}  \,.
\end{equation}
Notice that the field equations are not gauge invariant but transform into a set of equivalent equations, therefore the solution space is. Moreover, as $\,\delta \psi^c \neq 0\,$, the transformation (\ref{NC9}) cannot be a Noether symmetry and the variation $\delta\mathcal{L}$ is not a total divergence of $W^b$ that fulfills the asymptotic condition (\ref{TD3}) discussed in Section \ref{S2.1a}.

It is well-known that gauge invariance imposes that the field equations must satisfy certain off-shell identities. As noted above, the current nonlocal Lagrangian lacks gauge invariance due to (\ref{NC12}), but it can be shown that such relations do exist even for this case. In particular,
\begin{equation}  \label{NC13}
D_c \psi^c \equiv 0   \,.
\end{equation}  
Indeed,\\[1ex]
$ \displaystyle{\hspace*{10em}  D_c D_a F^{ac} = \frac12\,[D_c,D_a] F^{ac} = - \frac{i}2\,[F_{ca},F^{ac}] = 0  \,.}$ \hfill $\Box$

\subsection{Application of the second Noether theorem  \label{S3.2}}
We have derived the relation (\ref{NC13}) by mere inspection; however, it could also be derived from the second Noether theorem (see \cite{Pons2011} for a review of the local case), in particular, from the identity (\ref{GG3a}).
Indeed, from (\ref{NC9}) and (\ref{GG1}), it follows that 
$$  R_a(x,y) = -\left(D_a \delta_x\right)_{(y)} \, $$
and, substituting this into (\ref{GG2b}) and including the properties (\ref{Ap10e}) and (\ref{Ap10i}), we easily reach
\begin{equation}  \label{NC14}
 N(x) = -\left(D_a \psi^a\right)_{(x)} \,.
\end{equation}
Next, using properties of the Moyal bracket (\ref{Ap10e}) and (\ref{Ap10f}) in equations (\ref{GG3}) and (\ref{GG3z}), we arrive at
\begin{equation}  \label{NC14m}
J^b(x) + K^b(x) = F^{ab}(x) \,D_a \varepsilon (x) + \psi^b(x)\, \varepsilon(x) - i V^b (x)\,,
\end{equation}
where
\begin{equation} \label{NC14a}
V^b(x) = \int_0^1 \D s\,\int_{\RR^4} \D z\,z^b\, f(\xi, y) \,, \qquad {\rm with} \qquad y = x+ s z\,, \qquad \xi = x+(s-1) z \,,
\end{equation}
and
\begin{equation} \label{NC14b}
f(\xi, y) = \left(D_a \varepsilon(y)\,F^{ac}(\xi) + \psi^c(\xi)\,\varepsilon(y) \right)\,\left[A_c,\delta_y\right]_{(\xi)}\,, 
\end{equation}
where (\ref{Ap10g}) has been included. From the definitions, it immediately follows that
$$ \frac{\D f(\xi, y)}{\D s} = z^b\,\left(\frac{\partial f}{\partial \xi^b} + \frac{\partial f}{\partial y^b} \right) =  z^b\,\frac{\partial f \left(x+(s-1)z, x+s z\right)}{\partial x^b} \,,$$
and using (\ref{NC14a}), we get
$$ \partial_b V^b(x) 
= \int_{\RR^4} \D z\,\int_0^1 \D s\,\frac{\D f(\xi, y)}{\D s} =  \int_{\RR^4} \D z\,\left[f(x,x+z)- f(x-z,x)\right] \,.$$
Recalling (\ref{NC14b}), this leads to
\begin{equation}  \label{NC14c}
\partial_b V^b = - F^{ab}\,[A_a,D_b\varepsilon] + \psi^a\,[A_a,\varepsilon] - D_b\varepsilon \,[A_a,F^{ab}] + \varepsilon\,[A_a,\psi^a] \,,
\end{equation}
where (\ref{Ap10b}) has been included. Now, using (\ref{NC14m}) and (\ref{NC14c}), we can represent the central term on the left-hand side of the Noether identity (\ref{GG3a}) as
\begin{eqnarray*}
\partial_b\left( K^b + J^b \right) & = &  \partial_b\left(\varepsilon\,\psi^b + F^{ab}\,\partial_a\varepsilon - i F^{ab}\,[A_a,\varepsilon] - i V^b \right) \\[1ex]
 & = & -i\,\partial_b\left(\varepsilon\,[A_a, F^{ab}] + F^{ab}\,[A_a, \varepsilon]\right) - i \partial_b V^b\,
\end{eqnarray*}
that, using Leibniz's rule, equations (\ref{NC6}) and (\ref{NC1a}), and the Jacobi identity (\ref{Ap9}), we obtain
\begin{equation}  \label{NC14d}
\partial_b(J^b + K^b) = \frac{i}2\,F^{ab}[F_{ab},\varepsilon]\,.
\end{equation} 
Finally, substituting this and (\ref{NC12}) into the Noether identity and using (\ref{NC14}), we arrive at 
 $$ \varepsilon(x)\,D_a \psi^a(x) \equiv 0 \,,$$
that is the offshell relation (\ref{NC13}).

\subsection{The First Noether theorem. The canonical energy-momentum tensor  \label{S3.4}}
We will now discuss the invariance of the Lagrangian density (\ref{NC5}) under infinitesimal Poincar\'e transformations
\begin{equation}  \label{NC20}
\delta x^b = \varepsilon^b+ \omega^b_{\;\,c} x^c \,, \qquad \delta A_b(x) = -\omega^a_{\;\,b} A_a(x) - A_{b|c}(x)\, \delta x^c \,,
\end{equation}
where $\omega^{ab}$ is skewsymmetric, i. e. $\omega^{ab} + \omega^{ba} = 0$. Since the nonlocal Lagrangian (\ref{NC5}) does not depend explicitly on $x$ and is invariant under infinitesimal Poincar\'e transformations, we have $W^b = 0$ in equation (\ref{A10a}). Consequently, the Noether conserved current (\ref{A10}) is 
$$ J^b(x) = -\frac14\,F^{ad}(x) \,F_{ad}(x)\,\delta x^b + \int_{\RR^4} \D z\,z^b \int_0^1 \D s\,\lambda^d(y-z,y)\,\times \left[-\omega^a_{\;\,d} A_a(y) - A_{d|c}(y)\, \delta y^c \right] \,, $$
with $\,y= x+s z\,$. Using now (\ref{NC7}) and the fact that $\,\delta y^c = \delta x^c + s\, \omega^c_{\;\,e} z^e\,$, we easily arrive at
\begin{equation}  \label{NC22}
J^b = \mathcal{T}_ c^{\;\,b}\,\delta x^c + \frac12\,\omega^{ce}\,\mathcal{S}_{ec}^{\;\;\;b}  \,,
\end{equation}
where
\begin{equation}  \label{NC22a}
\mathcal{T}_c^{\;\,b}(x) :=  -\frac14\,F^{ad}(x) \,F_{ad}(x)\,\delta_c^b - \int_{\RR^4} \D z\,z^b \int_0^1 \D s\,F^{da}(y-z)\,\left( D_a \delta_y\right)_{(y-z)}\,A_{d|c}(y) 
\end{equation}
is the canonical energy-momentum tensor (or  current), and 
\begin{equation} 
\frac12\,\mathcal{S}_{ec}^{\;\;\;b}(x) :=  - \int_{\RR^4} \D z\,z^b \int_0^1 \D s\,F^{da}(y- z)\,\left( D_a \delta_y\right)_{(y-z)}\,\times 
\left[ \eta_{d[e} A_{c]}(y) + s\,z_{[e}\,A_{d|c]}(y) \right]
\end{equation} 
is the internal angular momentum current, which captures both the pure spin and a contribution from nonlocality.  

Now, the current (\ref{NC22}) depends on ten independent constant parameters, namely $\varepsilon^b$ and $\omega^{ce}$, and its local conservation implies the local conservation of both the canonical energy-momentum $\mathcal{T}_c^{\;\,b}$ and the total angular momentum current $\mathcal{J}_{ec}^{\;\;\;b}$ separately, i. e.
$$ \partial_b \mathcal{T}_c^{\;\,b} = 0 \qquad {\rm and} \qquad \partial_b \mathcal{J}_{ec}^{\;\;\;b} = 0 \,, $$
where 
\begin{equation}  \label{NC22c}
\mathcal{J}_{ec}^{\;\;\;b}:= 2 x_{[e}\mathcal{T}_{c]}^{\;\,b} +\mathcal{S}_{ec}^{\;\;\;b} \,
\end{equation}
splits in an orbital part and an internal part.

Now, to simplify the expressions of the canonical energy-momentum tensor and the internal angular momentum current, we can use the identity (\ref{Ap10i}) in Appendix A
$$  \left(D_a\delta_y\right)_{(x)} = - \left(D_a\delta_x\right)_{(y)} = \delta_{|a}(x-y)- i\,[A_a,\delta_y]_{(x)} \,.$$
This allows us to express them as 
\begin{equation} \label{NC23a}
\mathcal{T}_c^{\;\,b}(x) = -\frac14\,F^{ad}(x)F_{ad}(x)\,\delta_c^b + F^{bd}(x)\,A_{d|c}(x) + i \int_0^1 \D s \int_{\RR^4} \D z \,z^b
F^{da}(y-z)\,A_{d|c}(y) [A_a,\delta_y]_{(y-z)}
\end{equation}  
and
\begin{equation} \label{NC23b}
\mathcal{S}_{ec}^{\;\;\;b}(x) = F^b_{\;[e}(x)\,A_{c]}(x) + i\,\int_0^1 \D s \int_{\RR^4} \D z\,z^b F^{da}(y- z)
\left[ \eta_{d[e} A_{c]}(y) + s\,z_{[e}\,A_{d|c]}(y) \right] \, [A_a,\delta_y]_{(y-z)} \,,
\end{equation}
where $\,y:= x+ s z\,,$ and
$$[A_a,\delta_y]_{(y-z)} =  \int_{\mathbb{R}^4} \mathrm{d}u\, A_a(y+u) \tilde M(u,z)\,.$$

The similarity between the initial terms on the right-hand side of both expressions and their homologous in the local theory is noteworthy. The integral terms, on the other hand, represent nonlocal corrections in integral form. Notice also that $\mathcal{T}^{cb}$ is not symmetric. Achieving symmetrization would require resorting to the construction of the Belinfante-Rosenfeld energy-momentum tensor \cite{BELINFANTE1940,rosenfeld1940}, as was done in \cite{Heredia1}, but this is beyond the scope of the present paper.

On the contrary, it is important to mention the exploration of an infinite series representation for the energy-momentum tensor, as discussed in \cite{Moeller2002,Luca2022}, offering an alternative perspective. However, it is also worth mentioning that in \cite{Heredia2020v1}, the summation of this kind of infinite series was undertaken to attain a more concise representation of the energy-momentum tensor. This summation was exemplified within the context of classical electrodynamics in dispersive media, and the result was an integral expression (quite close) to the one presented in this manuscript (or in \cite{Heredia2022, Heredia_PhD}).

\section{The dynamic space. Solving the field equations  \label{S3.1a}}
Seeking a suitable parametrization for the dynamic space $\mathcal{D}\,$, i. e., the class of solutions of the field equations (\ref{NC8}), we write them as
\begin{equation}  \label{NC8a}
 \psi^c(x) \equiv \mathcal{E}^c_a  A^a + \Phi^c (A_a)  = 0 \,,
\end{equation}
with
\begin{equation}  \label{SWM1}
\mathcal{E}^c_a := \Box \,\delta^c_a - \partial^c \partial_a  \qquad {\rm and} \qquad 
\Phi^c (A_a) := - i\,\partial_a \left[A^a,A^c\right] - i\,\left[A_a,F^{ac}\right] \,.
\end{equation} 
We then define
\begin{equation}  \label{SWM2}
Z_a := A_a + \Box^{-1} \Phi_a(A_e) \,,
\end{equation} 
where $\Box^{-1}$ means a chosen particular solution of the D'Alembert inhomogeneous equation. We easily obtain that
$$ \mathcal{E}^c_a Z^a \equiv \psi^c - \Box^{-1} \partial^c\left(\partial_a \Phi^a\right) \,. $$
Now, from the identity (\ref{NC13}) and the fact that $\partial_c \mathcal{E}^c_a \equiv 0$, we get that
\begin{equation}  
\partial_a \Phi^a \equiv i\, \left[A_a,\psi^a\right]\,, 
\end{equation} 
and, therefore, 
\begin{equation}  \label{SWM3}
\mathcal{E}^c_a Z^a \equiv \psi^c - i\,\Box^{-1}\partial^c \left[A_a,\psi^a\right]  \,.
\end{equation}
Whence it follows that $\,\mathcal{E}^c_a Z^a= 0\,$ whenever $\,\psi^c = 0\,$.

Thus the definition (\ref{SWM2}) transforms the nonlocal system $\,\psi^c = 0\,$ into a local one $\,\mathcal{E}^c_a Z^a= 0\,$, which is nothing but Maxwell equations. Were it invertible, we could obtain the general solution of the nonlocal field equations (\ref{NC8}) as the inverse image of the general solution of the Maxwell field equations.

A way to clarify under what conditions it is possible to reverse the relation (\ref{SWM2}) could be to take it as a $\mathcal{C}^1$ map, $\; A_a \rightarrow Z_a\,$, connecting two Banach spaces. Since $A_a= 0$ is mapped into $Z_a= 0$ and the Jacobian map,
$$ \left.\frac{\delta Z_a(x)}{\delta A_c(y)}\right|_{A=0} = \delta_a^c \,\delta(x-y) \,,$$
is obviously an isomorphism, then the inverse function theorem \cite{Choquet1982} 
assures that the inverse map, $\, A_a = A_a(Z_b)\,$,  exists and is $\mathcal{C}^1$ in some respective neighbourhoods of $A_a= 0$ and $Z_a = 0\,$. However, working it out in detail is beyond the scope of the present paper.

This argumentation ensures the existence and uniqueness of perturbative solutions, i. e. in a neighbourhood of $A_a= 0$, but says nothing about other regions in the dynamic space.

Now, given a solution $Z^a$ of the Maxwell equations $\,\mathcal{E}^c_a Z^a=0 \,$, let us consider the field $A_a(Z)$ derived by solving (\ref{SWM2}) in that neighbourhood of $Z_a=0$. Is it a solution of the nonlocal equations $\,\psi^c(A) = 0\,$?

From equation (\ref{SWM3}) and the fact that $Z_a$ is a solution we have that
$$   \psi^c(x)  - i\,\Box^{-1}\partial^c \left[A_a,\psi^a  \right]_{(x)} = 0 \,, $$
which is a homogeneous linear equation on $\psi^a$. The Jacobian map (with respect to $\psi^a(y)$),
$$ \delta_a^c \,\delta(x-y)- i\,\Box^{-1}\,\partial^c \left[A_a,\delta_y  \right]_{(x)}\,, $$ 
is non-degenerate, provided that $A_a$ is small enough (in the sense of a suitable norm). Therefore, the homogeneous linear equation admits only the trivial solution in a neighbourhood of $A_a=0\,$.

\subsection{Connecting the non-commutative nonlocal gauge with a commutative local one }
The definition (\ref{SWM2}) maps the dynamic space $\mathcal{D}$ of the nonlocal field $\,A_a\,$ onto the dynamic space $\mathcal{D}_M$ of the Maxwell field $Z_a\,$. This local commutative gauge theory acts as a local alias for our nonlocal non-commutative theory. Let us see what the nonlocal gauge transformation $\,\delta A_a = D_a\varepsilon\,$ becomes in terms of the local field $Z_a\,$. 

To begin with, (\ref{SWM2}) implies that $\,\delta Z_a= \delta A_a + \Box^{-1} \delta \Phi_a\,$ and from the definition (\ref{SWM1}) it follows that
\begin{equation} \label{SWM4}
i\,\delta \Phi^a = \partial_c\left(\left[D^c\varepsilon,A^a\right] - \left[D^a\varepsilon,A^c\right]\right) + \left[D_c\varepsilon,F^{ca}\right]- i\left[A_c,\left[F^{ca},\varepsilon\right]\right] \,,
\end{equation}
where (\ref{NC10}) has been included. Now, after a little algebra, we obtain that
$$ \left[D^c\varepsilon,A^a\right] - \left[D^a\varepsilon,A^c\right] = \partial^c\,\left[\varepsilon,A^a\right] - \partial^a\left[\varepsilon,A^c\right] - \left[\varepsilon, F^{ca}\right] $$
and that 
$$ \left[D_c\varepsilon,F^{ca}\right]- i\left[A_c,\left[F^{ca},\varepsilon\right]\right] = \partial_c\left[\varepsilon, F^{ca}\right] \,,$$
where we have used that $\,\psi^a = 0\,$ because $\,A_a\,$ belongs to the dynamic space $\mathcal{D}$. Substituting these in (\ref{SWM4}) we obtain
$$ \delta \Phi^a = - i\,\Box \left[\varepsilon,A^a\right] + i\,\partial^a \partial_c \left[\varepsilon,A^c\right]  \,,$$
which, combined with (\ref{SWM2}) and $\,\delta A_a = D_a\varepsilon\,$, leads to
$\; \delta Z_a = D_a \varepsilon + i\,\left[A_a,\varepsilon\right] + i \,\partial_a \partial_c \Box^{-1}\,\left[\varepsilon, A^c\right]  \;$,
that is
\begin{equation}  \label{SWM5}
\delta Z_a = \partial_a f \,, \qquad {\rm with} \qquad   f= \varepsilon + i \,\Box^{-1} \partial_c\left[\varepsilon, A^c\right] \,.
\end{equation}

Since $f$ depends on $A_a$ as well as $\varepsilon$, this relation does not define a homomorphism group, and it could not be otherwise because one gauge group is not abelian and the other is. 

We have managed to find a relationship between the two Gauge transformations at the level of the dynamic space. This proposal is different from the one of Seiberg and Witten \cite{Seiberg1999} which establishes, only to the first order of approximation in the series of derivatives formalism, a correspondence between the non-commutative gauge and the commutative gauge in the kinematic space.

\subsection{The solutions}
The system of equations
\begin{equation}  \label{SWM6}
\mathcal{E}^c_a Z^a := \Box \,Z^c - \partial^c \partial_a Z^a = 0
\end{equation} 
is linear and the given of a set of initial data does not determine a unique solution. As a matter of fact, the equations are fulfilled by $\,\partial^a f\,$ for any arbitrary function $f$. In order to select a unique solution some gauge fixations are necessary. Here we shall use the Lorenz gauge condition plus a supplementary condition to fix the second kind gauge, namely
\begin{equation}  \label{SWM7}
\partial_a Z^a = 0  \qquad {\rm and} \qquad Z^4 = 0\,.
\end{equation} 
This  reduces the problem to solving simultaneously 
\begin{equation}  \label{SWM7a}
\Box Z^c = 0 \,, \qquad \quad\partial_a Z^a = 0 \,, \qquad \quad Z^4 = 0\,.
\end{equation} 
For the sake of convenience we will work with the Fourier transform (FT),
\begin{equation} \label{SWM8}
\hat Z^a(k)= \frac1{(2\pi)^2}\,\int_{\RR^4}\D x\,e^{-i k x}\, Z^a(x)   \,,
\end{equation}
in terms of which the field equations and gauge fixings respectively read
\begin{equation} \label{SWM8a}
k^b k_b \,\hat Z_c(k)= 0  \,, \qquad \quad k^a \hat Z_a = 0 \,, \qquad \quad \hat Z_4 = 0\,, 
\end{equation}
whose general solution is \cite{Vladimirov}
\begin{equation}  \label{e108z}
 \hat Z_c(k^e) := \sum_{\nu=\pm} W_{\nu c}(\mk)\,\delta(k^4 - \nu |\mk|) \,, 
\end{equation}
where the four-vectors $\,W_{\nu c}(\mk)\,$ are functions of $\mk$, the spatial part of $\, k_\nu^e :=(\mk,\,\nu |\mk|)\, , \quad \nu =\pm $. Furthermore, the gauge fixings imply that each of these four-vectors can be written in terms of a polarization vector $\mU_\mu(\mk)$ as
\begin{equation} \label{e114v}
 W_{\mu\,c}(\mk) = \left(\mU_\mu(\mk), \, 0 \right)  \,, \qquad {\rm with} \qquad \mk \cdot\mU_\mu (\mk) = 0  \,.
\end{equation}

Similarly as in the case of Maxwell field, the polarization vectors $\,W_{\pm\,a}(\mk)\,$ are connected with the initial data for the field $Z_a\,$. Indeed, the spatial FT of the initial data $\,Z_a(\mx,0)\,$ and  $\,\partial_t Z_a(\mx,0)\,$ are
$$\mathcal{F}_3\left(Z_a(\mathbf{x},0)\right)=\frac1{\sqrt{2 \pi}}\,\int_\RR \D k^4\,\hat Z_a(\mk, k^4) =  W_{+\,a}(\mk) +  W_{-\,a}(\mk) $$
and 
$$\mathcal{F}_3\left(\partial_t Z_a(\mathbf{x},0)\right)= \frac{i}{\sqrt{2 \pi}}\,\int_\RR \D k^4\, k^4\,\hat Z_a(\mk, k^4) =  i\,|\mk|\,\left(W_{+\,a}(\mk) -  W_{-\,a}(\mk) \right) \,, $$
where (\ref{e108z}) has been included. Then it easily follows that
$$  W_{\pm\,a}(\mk) = \frac{1}{2} \mathcal{F}_3\left(Z_a(\mathbf{x},0)\right) \mp \frac{i}{2|\mathbf{k}|} \mathcal{F}_3\left(\partial_t Z_a(\mathbf{x},0)\right)\,.$$ 

Due to the gauge fixations (\ref{SWM7}), the solutions (\ref{e108z}-\ref{e114v}) do not exhaust the entire dynamic space $\mathcal{D}_M$. Indeed, the gauge transformation $Z^{\prime}_a = Z_a +\partial_a f\,$ define an equivalence relation among the solutions in $\mathcal{D}_M$
and each solution (\ref{e108z}-\ref{e114v}) is a representative of its equivalence class in the quotient space $ \mathcal{D}_M^{\rm ph}=\mathcal{D}_M / {\rm gauge}\,$, which acts as the physical dynamic space for the alias local theory.

The general solution of the nonlocal set of equations (\ref{NC8a}) is then obtained by solving the implicit functional equation
\begin{equation}  \label{e108}
 \hat A_c(k^e) =  \frac1{k^a k_a}\, \hat \Phi_c(k^e) +  \hat Z_c(k^e) \,,
\end{equation}
where $\, \hat \Phi_c\,$ is the FT of $\,\Phi^c \,$ given by (\ref{SWM1}), that is
$$ \Phi^c := - i\,\left[\partial_a A^a, A^c\right] - i\,\left[A_a, 2 \partial^a A^c - \partial^c A^a  \right] - \left[A_a, \left[ A^a , A^c \right] \right]\,.$$
As it can be easily checked, the Lorenz gauge condition $\partial_a Z^a = 0 $ implies that $\partial_a A^a = 0 $, with which we arrive at 
\begin{equation}  \label{e108a}
 \Phi^c :=  - i\,\left[A_a, 2 \partial^a A^c - \partial^c A^a  \right] - \left[A_a, \left[ A^a , A^c \right] \right]\,,
\end{equation}
and its FT is
\begin{equation}  \label{e108b}
 \hat \Phi^c :=  \left[\hat A_a, 2 \,k^a \hat A^c - k^c \hat A^a \right] - \left[\hat A_a, \left[\hat A^a , \hat A^c \right] \right]\,,
\end{equation}
where (\ref{Ap80}) has been used.

Equation (\ref{e108}) can be solved perturbatively by assuming that the more Moyal brackets a term contains, the smaller it is. As $\,\hat \Phi_c\,$ is the only term containing Moyal brackets,  $\,\hat A_c(k^e) = \hat Z_c(k^e) + O(1)\,$ is the seed on which the perturbative solution is built. As commented before, a more rigorous proof of the existence of one solution $\hat A_c$ for each given $\hat Z_c$ might be approached on the basis of the inverse function theorem on a Banach space \cite{Choquet1982}.

The solution up to the second order is
$$ \hat A_a = \hat Z_a + \frac1{k^b k_b}\,\left[\hat Z^c, 2 \,k_c \hat Z_a - k_a \hat Z_c \right] + O(2)\,,$$
which, using (\ref{e108z}), leads to 
\begin{eqnarray}
\hat A^a(k) & = & \hat Z^a(k) + \frac{i}{2\pi^2}\,\sum_{\nu,\mu= \pm } \int_{\RR^6} \D \mq\,\D \Mp\, W_{\nu c}(\mq)\,W_{\mu b}(\Mp)\,\delta(\mq+\Mp-\mk)\,\delta(k^4 -\nu |\mq| - \mu |\Mp|)\,
\nonumber\\[2ex] \label{m11}
 & & \times \sin\left(\frac{q_\nu \theta p_\mu}2\right) \, h^{acb}(q_\nu,p_\mu)+ O(2)
\end{eqnarray}
with
\begin{equation} \label{m11a}
h^{acb}(q_\nu,p_\mu) := \frac1{|\mk|^2 - (\nu |\mq| +\mu |\Mp)^2}\,\left(p_{\mu}^c \eta^{ab} - q_{\nu}^b \eta^{ac} -\frac12\eta^{cb} (p_\mu^a - q_\nu^a) \right)\,,
\end{equation}
and $\,p_\mu^a := (\Mp, \mu\,|\Mp|)\,, \quad \mu= \pm 1\,$.

%% file: NC1-Hamiltonian.tex
\section{Hamiltonian formalism on the physical dynamic space \label{S3.5}}
We will next establish a Hamiltonian formalism on $\mathcal{D}^{\rm ph}\subset\mathcal{K},$ by employing the methodologies introduced in \cite{Heredia2022} and summarized in Section \ref{SS3.2}.  We choose to set up 
the Hamiltonian formalism on $\mathcal{D}^{\rm ph}$ through symplectic techniques rather than Dirac brackets because the former allow a more straightforward implementation of constraints. 

According to what is explained and justified in \cite{Heredia2022}, the physical dynamic space is the submanifold of the class of functions (that are  differentiable as many times as necessary) such that they fulfill the Euler-Lagrange equations (\ref{NC8}) taken as constraints. The Hamiltonian formalism is then based on the presymplectic form (\ref{eq:SF}):
\begin{equation}   \label{Sp1a}
 \omega(A) = \int_{\RR^4} \D y\,\delta P^a(y, A) \wedge \delta A_a(y) \,,
\end{equation}
where the canonical momenta (\ref{AM3}) are
\begin{equation}   \label{Sp1c}
 P^a(y, A):=  \int_{\RR^4} \D z\,\left[\Theta(y^4)-\Theta(y^4-z^4)\right]\,\lambda^a(A,y-z,y) \,,
\end{equation}
$\lambda^a$ is defined by (\ref{L2o}), and the dependence on $A_b$ is of functional type. In the language of \cite{Heredia2022}, we refer to this as the momentum constraints $\, p(y) = P(y,  A)\,.$ Up to this point, we shall use the plural form for canonical momenta (or momentum constraints) to refer to one equation since it represents a continuous infinity of constraints, one for each $y\in \RR^4$. Furthermore, we will write $P^a(y)$ instead of $P^a(y,A)$ to simplify the notation.

The Hamiltonian dynamics is then ruled by the Hamiltonian function (\ref{AM4}):
\begin{equation}   \label{Sp1}
 h(A) = - L(A) + \int_{\RR^4} \D y\, P^a(y)\,\partial_4 A_a(y) \,,
\end{equation}  
where 
\begin{equation}   \label{Sp1d}
 L(A):=  \int_{\RR^3} \D \mathbf{y}\, \mathcal{L}\left(T_\mathbf{y}A\right)  
\end{equation}    
is the Lagrangian.
Furthermore, the field $A_b$ has to be expressed in terms of a suitable parametrization of $\mathcal{D}^{\rm ph}\,$, e. g. the parameters $\mU_\pm(\mk)\,$ in (\ref{m11}). Once this is set, we shall be able to ascertain whether $\omega$ is non-degenerate, hence symplectic, or not.

Using (\ref{NC7}) and introducing the variable $\zeta = y - z\,$, the momentum constraints become
$$  P^a(y) = \int_{\RR^4} \D \zeta\, \left[\Theta(y^4)-\Theta(\zeta^4)\right]\,F^{ac}(\zeta)\, \left(D_c \delta_y\right)_{(\zeta)}\,,$$
and, including (\ref{Ap10k}), we have that 
$\; P^a(y) = - D_c\left\{\left(\Theta(y^4)-\Theta(\zeta^4)\right)\,F^{ac}(\zeta)\right\}_{(\zeta=y)} \;$ 
or equivalently
\begin{equation}  \label{H7z} 
P^a(y) = -\Theta(y^4)\,\psi^a(y) + D_c\left(\Theta(y^4)\,F^{ac}(y)\right) \,.
\end{equation}
Developing it a bit further, we arrive at
\begin{equation}  \label{H7} 
P^a(y) = \delta(y^4)\,F^{a4}(y) + i\,\left[\Theta(y^4)\,F^{ac}(y), A_c(y)\right] - i\,\Theta(y^4)\,\left[F^{ac}(y), A_c(y)\right] \,
\end{equation}
and, combining this with (\ref{Sp1}) and (\ref{Sp1d}), we obtain that $\; h =  \int_{\RR^3} \D \mathbf{y}\,\mathcal{H} \,$, where
\begin{equation}  \label{NC11a}
 \mathcal{H} = - \mathcal{L} + A_{a|4}(\mathbf{y},0)\,F^{a4}(\mathbf{y},0) + i \,\int_{\RR} \D y^4\,\left(\left[\Theta(y^4)\,F^{ac}(y), A_c(y)\right] - \Theta(y^4)\,\left[F^{ac}(y), A_c(y)\right] \right)
\end{equation}
is the Hamiltonian density.

For the sake of convenience 
the Fourier transforms of the potentials and the momenta will be useful, namely
\begin{equation} \label{M1}
\hat{A}_a(k) = \frac1{(2\pi)^2} \,\int_{\RR^4} \D y\,e^{-i k y}\,A_a(y)  \qquad {\rm and} \qquad 
\hat{P}^a(k) = \frac1{(2\pi)^2} \,\int_{\RR^4} \D y\,e^{-i k y}\,P^a(y)\,. 
\end{equation}
In terms of these variables the (pre)symplectic form (\ref{Sp1a}) on the dynamic space is
\begin{equation}   \label{Sp1b}
 \omega = \int_{\RR^4} \D k\,\delta \hat{P}^a(k) \wedge \delta \hat{A}_a(-k) \,.
\end{equation}

\subsection{The momenta}
The Fourier transform of the momenta (\ref{H7}) is
\begin{equation} \label{M2}
\hat P^a(k)= \frac1{2\pi}\,\int_{\RR}\D\rho\,\hat{F}^{a4}(\mathbf{k},\rho) + i\,\mathcal{F}\left[\Theta(y^4)\,F^{ac}(y), A_c(y)\right]_{(k)} 
 - i\,\mathcal{F}\left(\Theta(y^4)\,\left[F^{ac}(y), A_c(y)\right]_{(k)}  \right) \,.
\end{equation}
Then, using the definition (\ref{Ap80}), we have that
$$ i\,\mathcal{F}\left[\Theta(y^4)\,F^{ac}(y), A_c(y)\right]_{(k)} = i\,\left[\mathcal{F}\left(\Theta(y^4)\,{F}^{ac}(y) \right)_{(k)}, \hat{A}_c(k)\right] \,$$
and, by the convolution theorem, 
$$  \mathcal{F}\left(\Theta(y^4)\,{F}^{ac}(y) \right)_{(k)} = \frac1{(2\pi)^2}\,\left(\mathcal{F}\Theta(y^4)\ast \hat{F}^{ac} \right)_{(k)} \,.$$
Therefore, using that $\;\displaystyle{\mathcal{F}\Theta(y^4)_{(k)} = \frac{2\pi i}{k^4 + i 0}\,\delta(\mk) }\,$, we arrive at
$$  \mathcal{F}\left(\Theta(y^4)\,{F}^{ac}(y) \right)_{(k)} = \frac{i}{2\pi}\,\int_{\RR^4} \D q\,\frac{\delta(\mq)}{q^4+i 0}\,\hat{F}^{ac}(k-q) \,,$$
which leads to
$$  i\,\mathcal{F}\left[\Theta(y^4)\,F^{ac}(y), A_c(y)\right]_{(k)} = -\frac1{2\pi}\int_{\RR} \frac{\D \tau}{\tau+i 0} \,\left[T_{-\tau}\hat{F}^{ac}, \hat{A}_c\right]_{(k)}  \,,$$
where $\,T_{-\tau}\hat{F}(k^b):= \hat{F}(k^b-\tau \,\delta_4^b)\,$, and similarly
$$  i\,\mathcal{F}\left(\Theta(y^4)\, \left[F^{ac}(y), A_c(y)\right]\right)_{(k)} = 
- \frac1{2\pi}\int_{\RR} \frac{\D \tau}{\tau+i 0} \,T_{-\tau}\left[\hat{F}^{ac}, \hat{A}_c\right]_{(k)}  \,.$$
Substituting the latter into (\ref{M2}), we finally obtain that 
\begin{equation} \label{M3}
\hat P^a(k)= \frac1{2\pi}\,\int_{\RR}\D\rho\,\hat{F}^{a4}(\mathbf{k},\rho)  -
\frac1{2\pi}\int_{\RR} \frac{\D \tau}{\tau+i 0} \,\left(\left[T_{-\tau}\hat{F}^{ac}, \hat{A}_c\right]_{(k)}- T_{-\tau}\,\left[\hat{F}^{ac}, \hat{A}_c\right]_{(k)} \right)  \,.
\end{equation}


\subsection{The symplectic form on the physical dynamic space }
The symplectic form $\omega \in \Lambda^2(\mathcal{D}^{\rm ph})$ results from replacing $\hat{P}^a(k)$ and $\hat{A}_a(-k)$ in (\ref{Sp1b}) with the constraints (\ref{M3}) and (\ref{e108}). Thus, substituting the lowest orders of these constraints, 
\begin{eqnarray} 
  \hat P^a(k) &=& \frac{i}{\pi}\,\sum_{\rho=\pm} k_\rho^{[a} W_\rho^{4]}(\mk) + O(1) \,,  \qquad  k_\rho^a := (\mk, \rho|\mk|) \label{P_C1}\\[1ex]
\hat A_c(-k) &=& \sum_{\nu=\pm} W_{\rho c}(-\mk)\,\delta(k^4+\nu |\mk|)  + O(1) \,, \label{A_C2}
\end{eqnarray} 
into the expression (\ref{Sp1b}) and keeping only the lowest order terms, we obtain that 
\begin{equation}  \label{sP2a}
\omega_0 =  \frac{i }\pi\, \int_{\RR^3}\D\mk\,|\mk|\, \delta_{jl} \delta U_{-}^j(\mk)\wedge \delta U_{+}^l(-\mk)\,,
\end{equation}
where (\ref{e114v}) has been included.

The derivation of the symplectic form to the next perturbative order is presented in detail in Appendix B,  and the result is
\begin{equation}  \label{sP3}
\omega =  \omega_0 + O(2)\,.
\end{equation}
Recall that the physical dynamic space $\,\mathcal{D}^{\rm ph}\,$ is the quotient of $\,\mathcal{D}\,$ over the full gauge equivalence, first and second kind, which is coordinated with $\,\mU_\pm(\mk)\,$, according to (\ref{e108z}-\ref{e114v}).

Seemingly the expression (\ref{sP3}) should hold at any perturbative order due to the relation (\ref{SWM2}) and the fact that the alias local theory is just Maxwell theory, whose symplectic form is exact.  
Equation (\ref{sP3}) also points out what the canonical coordinates are to this approximation. 
Since $Z_c(x)$ has to be real, eq. (\ref{M1}) implies that $\hat{Z}^{*}_c(k) = \hat{Z}_c(-k)\,,$, whence it easily follows that 
\begin{equation}\label{NC_canVar}
\mU^\ast_{\rho}(\mathbf{k}) = \mU_{-\rho}(-\mathbf{k})\,, \qquad \rho = \pm \,.
\end{equation} 
Substituting these in (\ref{sP3}), we obtain
\begin{equation*}
\omega = - \frac{i}\pi\, \int_{\RR^3}\D\mk\,|\mk|\, \delta U^l_{+}(\mk)\wedge \delta U^{\ast}_{+ l}(\mk) + O(2)\,,
\end{equation*} 
and, introducing the new variables 
\begin{equation}\label{NC_canVar_2}
 \mU(\mk) := \sqrt{\frac{|\mk|}{\pi}} \mU_{+}(\mk) \qquad \mbox{and its complex conjugate} \qquad \mU^\dagger(\mk) := \sqrt{\frac{|\mk|}{\pi}} \mU^\ast_{+}(\mk)\,,
\end{equation}
the symplectic form on $\,\mathcal{D}^{\rm ph}\,$ reduces to 
\begin{equation}
\omega = -i \,\int_{\RR^3}\D\mk\, \delta U^\dagger_l(\mk)\wedge \delta U^l(\mk) + O(2)\,.   
\end{equation}
It is thus evident that the modes $U_j(\mk)$ and $U^\dagger_l(\mk)\,$ are a set of canonical variables whose non-vanishing elementary Poisson brackets are
\begin{equation}\label{S5.3CQ}
   \left\{U_l(\mk), U^\dagger_j (\mk^\prime)\right\} = \delta_{lj}\, \delta (\mk-\mk^\prime) + O(2) \,.  
\end{equation}

\subsection{The Hamiltonian on the physical dynamic space}
As discussed at the beginning of Section \ref{S3.5}, the Hamiltonian $h$ defined on the dynamic space arises from implementing the constraints ---namely, canonical momenta and field equations--- into the Hamiltonian $H$ defined on the kinematic space. 

As we have done for the symplectic form, it is convenient to work in Fourier space, therefore, the Hamiltonian $h$ on $\mathcal{D}^{\rm ph}$ is
\begin{equation}\label{S54_h_D}
    h = i \int_{\mathbb{R}^4}\mathrm{d}k\, k^4\, \hat P^a(k) \hat A_a(-k) + \frac{1}{8\pi} \int_{\mathbb{R}^5}\mathrm{d}k\,\mathrm{d}\sigma\, \hat F^{ab}(k) \hat F_{ab}(-\mathbf{k},-\sigma)\,.
\end{equation}
Taking into account that at the lowest order  ---see Appendix B--- 
$$\hat F^{ab}(k) = 2 i k^{[a} \hat Z^{b]} + O(1) $$  
and substituting the lowest orders of the constraints (\ref{A_C2}-\ref{P_C1}) into the Hamiltonian (\ref{S54_h_D}), the latter becomes (after a bit of algebra) 
\begin{equation}  \label{S54_h_E}
    h_0 = \frac{1}{\pi} \int_{\mathbb{R}^3} \mathrm{d}\mathbf{k}\,|\mathbf{k}|^2\,\mU_{-}(\mk)\cdot \mU_{+}(-\mk)\,,
\end{equation}
where equation (\ref{e114v}) has been used. In Appendix B we explicitly derive the next perturbative order of the Hamiltonian and the result is null, that is,
\begin{equation}
    h = h_0 + O(2)\,.
\end{equation}
In terms of the canonical variables (\ref{S5.3CQ}) the Hamiltonian reads
\begin{equation}  \label{S54_h_F}
    h = \int_{\RR^3}\mathrm{d}\mk\,|\mk|\,\mU(\mk) \cdot \mU^\dagger(\mk)  + O(2)\,.
\end{equation}

As commented before for the symplectic form, the vanishing of the $O(1)$ terms in the Hamiltonian probably persists for higher orders and the relation (\ref{S54_h_E}) is likely exact. Indeed, taking $\mU_\pm(\mk)$ as coordinates of the dynamic space $\mathcal{D}^{\rm ph}\,$, they are the same as those for the local field $\hat Z_a$, which is nothing but a free Maxwell field. In fact, the implicit equation (\ref{SWM2}) establishes a  one-to-one correspondence between the nonlocal field $A_a$ and its alias, the free Maxwell field $Z_a$. 

Therefore, the nonlocal field $\hat A_a$ is nothing but a fashionable deformation of its local alias, so it might seem appropriated to set the nonlocal $\hat A_a$ aside and stick with the local $\hat Z_a$. We will proceed this way in what concerns the Hamiltonian formalism, canonical coordinates and quantization.

However, the nonlocal field comes to fore when the coupling to other fields is considered. For this reason, it is worth writing $\hat A^c$ in terms of the canonical variables $\mU$ and $\mU^\dagger$ and we have that ---see equation (\ref{Pots}) Appendix B for details---
\begin{eqnarray}  
\hat A^a(k) &=& \hat Z^a(k) +  \frac{i}{2\pi} \,\int_{\RR^6}\frac{\D\mq\, \D\Mp}{\sqrt{|\mq|\,|\Mp|}}\,\sin\left(\frac{q_+\theta p_+}{2}\right) \,\left\{ 2\,\delta(k^b-q^b_+ + p^b_+)\,S^a(\mq,\Mp)  + \right. \nonumber \\[2ex]  \label{Pots}
 & & \left. \delta(k^b-q^b_+- p^b_+)\,R^a(\mq,\Mp) - \delta(k^b+q^b_+ + p^b_+)\, R^{\dagger\,a}(\mq,\Mp)   \right\}  +O(2)  \,,
\end{eqnarray}
with
$$ \hat{\mathbf{Z}} = \sqrt{\frac{\pi}{|\mk|}}\,\left\{\mU(\mk)\,\delta(k^4-|\mk|) + \mU^\dagger(\mk)\,\delta(k^4+|\mk|) \right\}  
\,, \qquad \hat Z^4 = 0 \,,$$
$$ R^a(\mq,\Mp):= \frac1{\mk^2-(|\mq|+|\Mp|)^2}\,\left[\Mp\cdot\mU(\mq)\,\delta^a_l\,U^l(\Mp) - \mq\cdot\mU(\Mp)\,\delta^a_l\,U^l(\mq) - \frac12\,\mU(\mq)\cdot\mU(\Mp)\,\left(p_+^a - q_+^a\right) \right] $$
and
$$ S^a(\mq,\Mp):= \frac{-1}{\mk^2-(|\mq|-|\Mp|)^2}\,\left[\Mp\cdot\mU(\mq)\,\delta^a_l\,U^{\dagger\,l}(\Mp) + \mq\cdot\mU^\dagger(\Mp)\,\delta^a_l\,U^l(\mq)- \frac12\,\mU(\mq)\cdot\mU^\dagger(\Mp)\,\left(p_+^a+q_+^a\right) \right] \,.$$

\subsection{Quantization}
The quantum commutation relations \cite{Dirac1964} trivially follow from (\ref{S5.3CQ}) and the non-vanishing ones are 
\begin{equation}\label{S5.4-QC}
  \left[\mathbb{U}_l(\mathbf{k}), \mathbb{U}^\dagger_j(\mathbf{k}^\prime)\right] = \hbar\, \delta_{lj}\, \delta(\mk-\mk^\prime)\,.
\end{equation}
The "Blackboard Bold" type emphasises that $\mathbb{U}_l(\mathbf{k})$ and $\mathbb{U}^\dagger_j(\mathbf{k})$ are the quantum counterparts of the classical variables $U_l(\mathbf{k})$ and $U^\dagger_j(\mathbf{k})$, and are the adjoint of each other.  Additionally, we can introduce the annihilation and creation operators, $a_l(\mk)$ and $a_l^{\dagger}(\mk)$ respectively, as
\begin{equation}\label{S5.4-LR}
a_l(\mk):= \frac{1}{\sqrt{\hbar}}\,\mathbb{U}_l(\mathbf{k})   \, , 
\end{equation}
and its adjoint, with 
\begin{equation}\label{S5.4-AR}
    \left[a_l(\mk),  a^\dagger_{j}(\mk^\prime)\right] =\delta_{lj} \delta(\mk-\mk^\prime)\,.
\end{equation}
The corresponding Hamiltonian operator is
\begin{equation}   \label{S54_h_Fq}
\mathbb{H} =  \int_{\RR^3}\mathrm{d}\mk\, \hbar\,\omega(\mk) \left(\mathbb{N}(\mk) + \frac{1}{2}\right)\,,    
\end{equation}
where we have used equations (\ref{S5.4-LR}-\ref{S5.4-AR}), $\omega(\mk):= |\mk|$, and 
\begin{equation}   \label{S54_N_q}
\mathbb{N}(\mk):=a^\dagger_l(\mk)\,a^l(\mk)
\end{equation}
is the number operator. Notice quickly that the Hamiltonian operator is the same as for the electromagnetic field, as we expected \cite{zeidler2008}.

In turn the quantum operator for the nonlocal field in terms of the creation and anihilation operators $a$ and $a^\dagger$ is
\begin{eqnarray}  
\hat{\mathbb{A}}^a(k) &=& \hat{\mathbb{Z}}^a(k) +  \frac{i}{2\pi} \,\int_{\RR^6}\frac{\D\mq\, \D\Mp}{\sqrt{|\mq|\,|\Mp|}}\,\sin\left(\frac{q_+\theta p_+}{2}\right) \,\left\{ 2\,\delta(k^b-q^b_+ + p^b_+)\,\mathbb{S}^a(\mq,\Mp)  + \right. \nonumber \\[2ex]  \label{PotsQu}
 & & \left. \delta(k^b-q^b_+- p^b_+)\,\mathbb{R}^a(\mq,\Mp) - \delta(k^b+q^b_+ + p^b_+)\, \mathbb{R}^{\dagger\,a}(\mq,\Mp)   \right\}  +O(2)  \,,
\end{eqnarray}
with
$$ \hat{\mathbb{Z}}^b = \sqrt{\frac{\pi \hbar}{|\mk|}}\,\delta^b_l\left\{a^l(\mk)\,\delta(k^4-|\mk|) + a^{\dagger\,l}(\mk)\,\delta(k^4+|\mk|) \right\}   \,,$$
$$ \mathbb{R}^a(\mq,\Mp):= \frac{\hbar}{\mk^2-(|\mq|+|\Mp|)^2}\,\left[\Mp\cdot \ma(\mq)\,\delta^a_l\,a^l(\Mp) - \mq\cdot\ma(\Mp)\,\delta^a_l\,a^l(\mq) - \frac12\,\ma(\mq)\cdot\ma(\Mp)\,\left(p_+^a - q_+^a\right) \right] $$
and
$$ \mathbb{S}^a(\mq,\Mp):= \frac{-\hbar}{\mk^2-(|\mq|-|\Mp|)^2}\,\left[\Mp\cdot\ma(\mq)\,\delta^a_l\,a^{\dagger\,l}(\Mp) + \mq\cdot\ma^\dagger(\Mp)\,\delta^a_l\,a^l(\mq)- \frac12\,\ma(\mq)\cdot\ma^\dagger(\Mp)\,\left(p_+^a+q_+^a\right) \right] \,.$$

%% file: NC1-Conclusion.tex
\section{Conclusions}\label{SConcl}
In this article, we provide an extensive overview of the nonlocal Lagrangian formalism proposed in \cite{Heredia2,Heredia2022}, including the extension of Noether's first theorem. Additionally, we extend Noether's second theorem applied for these Lagrangians, highlighting their practical use in non-commutative U(1) --NCU(1)-- gauge theory.

The nonlocal nature of NCU(1) gauge theory is embodied by the Moyal $\ast$ operator, treatable as an infinite series or integral operator. In this article, we opt for the latter, motivated by the observations made in \cite{Heredia2023_1,Carlsson2016}, where it is discussed that the functional space of this operator might be wider than the infinite series one. In this context, we have applied the nonlocal Lagrangian formalism and obtained its field equations (\ref{NC8}). Next, we have studied how these field equations are transformed through the nonlocal gauge transformation (\ref{NC9}). We have observed that it leaves the solution space invariant but not the field equations themselves. They are transformed into equivalent ones. For this reason, we have concluded that the nonlocal gauge transformation cannot be a Noether symmetry, otherwise the field equations would be completely invariant. 

Upon close examination, we have observed that the identity (\ref{NC13}) holds true even though the nonlocal Lagrangian is not gauge invariant. Remarkably, this identity has been corroborated by the extension of Noether's second theorem for nonlocal Lagrangians, thus affirming the validity of it. It is worth noting that, in this specific case, the use of the developed extension might have been deemed unnecessary, as mentioned earlier, a simple inspection reveals it. However, for more intricate scenarios, this extension can prove exceptionally valuable.

Since the NCU(1) gauge theory does not depend explicitly on the point $x^b$ and its Lagrangian density is Poincar\'e invariant, we have applied Noether's first theorem to obtain the canonical momentum energy tensor and the internal angular momentum current in integral form for this symmetry. Both exhibit a local part and a nonlocal part coming from nonlocality hidden in $F_{ab}$. 

One of the strengths of the nonlocal Lagrangian formalism was to be able to propose a Legendre transform in order to construct a Hamiltonian formalism for nonlocal theories. We have applied this Hamiltonian formalism to the NCU(1) gauge theory. To do so, we have studied the space of solutions, and we have proposed the construction of a perturbative solution through the equation (\ref{e108}) which, for simplicity of calculation, we have kept up to second order; the mathematical foundation of this proposal has been the inverse function theorem in Banach spaces. Once the pertubative solution is obtained, we have introduced it in the momenta constraint (\ref{M3}), and with them, we have calculated the symplectic form and the Hamiltonian. Furthermore, we have obtained the canonical variables and the elementary Poisson brackets, and implemented the canonical quantization. We have observed that up to second order terms the symplectic form and the Hamiltonian coincide with the electromagnetic theory, i.e. the first order of both vanish. This result is not surprising since equation (\ref{e108}) establishes a (pertubative) one-to-one correspondence between the nonlocal field $A^a$ and the local field $Z^a$. For this reason, and in the light of this result, we suggest that this theory (which is nothing more than a fashionable deformation of Maxwell's electromagnetic theory) be set aside and work with the local theory. It is important to note that, if this theory were coupled with another field, the latter would not be true, and one would have to work with the nonlocal theory since the coupling would manifest itself in the context of $A^a$ and not $Z^a$. 

Finally, we have reviewed the Seiberg-Witten map. We have observed that the mapping (96) partially succeeds in connecting the nonlocal and nonabelian gauge theory with a local and abelian theory. We emphasize "partially" because the goal is only achieved at the level of the dynamical space, and not in the kinematic space which is where (at first order of $\theta^{ab}$) the Seiberg-Witten map connects them. 

Currently, the focus in the field of non-locality lies predominantly in the field of nonlocal gravity. It would be very interesting to be able to study this nonlocal formalism, together with Noether's second theorem, in models of nonlocal gravity \cite{Biswas2012,Calcagni2018,Kol2021,Capozziello2022_NS} and to be able to see what conclusions can be drawn from it. Complex but challenging. 

\section*{Acknowledgment}
Funding for this work was partially provided by the Spanish MINCIU and ERDF (project ref. RTI2018-098117-B-C22) and by the Spanish MCIN (project ref. PID2021-123879OB-C22).

%% file: NC1-Appendix.tex
\section*{Appendix A: The Moyal $\ast$ product and bracket} 
The Moyal $\ast$ product introduced in (\ref{NC2}) is to be understood as the Taylor expansion of the exponential; this yields an infinite series which, using the  inverse Fourier transform, 
$$ f(x+\alpha) = \frac1{(2\pi)^2} \,\int_{\RR^4} \D q\,e^{i q (x+\alpha)}\,\hat f(q)\,, \qquad  $$
can be translated into an integral expression ---$\hat{f}(q)$ is the Fourier transform of $f(x)$---, and similarly for $g(x+\beta)\,$.
Indeed, 
\begin{eqnarray*}   
(f\ast g )_{(x)} &:=& \left[\exp\left(\frac{i}2\,\theta^{ab}\,\frac{\partial \;}{\partial \alpha^a}\,\frac{\partial \;}{\partial \beta^b}\right) \,f(x+\alpha)\,g(x+\beta)\right]_{\alpha=\beta=0} \\
 & = & \frac1{(2\pi)^4} \,\int_{\RR^8} \D q\,\D k\, \hat f(q) \,\hat g(k)\,\exp\left(-\frac{i}{2}\,\theta^{ab} q_a k_b + i\,q_a x^a + i\,k_a x^a \right)\,, 
\end{eqnarray*}
and, substituting the expressions for the Fourier transforms $\hat f(q)$ and $\hat g(k)$, we arrive at
\begin{equation}  \label{Ap1}
(f\ast g )_{(x)} = \int_{\RR^8} \D y\,\D z\,f(y)\,g(z)\,M(x-y,x-z) \,,
\end{equation}
with
\begin{equation}  \label{Ap2}
M(y,z) := \frac1{(2\pi)^8}\,\int_{\RR^8} \D q\,\D k\, \exp\left(-\frac{i}{2}\,\theta^{ab} q_a k_b  + i q_a y^a + i k_b z^b  \right)\,. 
\end{equation}
Defining $\tilde k^a := \theta^{ab} k_b \,$, the above expression becomes
\begin{equation}  \label{Ap3}  
M(y,z) :=\frac1{(2\pi)^8}\,\int_{\RR^8}\D q\,\D k\,e^{-\frac{i}2 \,q_a(\tilde k^a -2 y^a)}\, e^{i k_b z^b} = \frac1{(2\pi)^4}\,\int_{\RR^4} \D k\, e^{i k_b z^b }\, \delta\left(\frac12 \,\tilde k^a - y^a\right) \,.
\end{equation}
For the sake of simplicity, we shall hereon assume that $\theta^{ab}$ is non-degenerate.  Therefore, $\tilde\theta_{bc}$ exists such that $ \,\theta^{ab} \tilde\theta_{bc} = \delta^a_c \,$
and, using that 
$$\delta\left(\frac12 \,\theta^{ab} k_b - y^a\right) = \frac{2^4}{|\theta|}\,\delta(k_b - 2\,\tilde\theta_{ba} y^a) \,, \qquad \quad |\theta| := \det(\theta^{ab})\,,$$ 
we finally arrive at
\begin{equation}  \label{Ap4}  
M(y,z) := \frac1{\pi^4 |\theta|}\, e^{2 i\,z\tilde\theta y} \,, \qquad {\rm where }\qquad  z\tilde\theta y = z^b\tilde\theta_{bc} y^c \,.
\end{equation}
Thus, the Moyal product (\ref{Ap1}) can be written as
\begin{equation}  \label{Ap7a}  
(f\ast g)_{(x)} = \frac{1}{\pi^4|\theta|}\,\int_{\RR^8} \D u\,\D v\,f(x-u)\,g(x-v)\, e^{2 i \,v \tilde\theta u} \,.
\end{equation}
The Moyal bracket, $\,[f,g] = f\ast g - g\ast f\,$, is
\begin{equation}  \label{Ap7}
[f,g]_{(x)} = \int_{\RR^8} \D u\,\D v\,f(x+u)\,g(x+v)\, S(u,v)  \,, \quad {\rm where} \quad S(u,v) := \frac{2 i}{\pi^4|\theta|}\,\sin(2 v \tilde\theta u)\,.  
\end{equation}
Similar expressions can be found for the degenerate cases.

\subsection*{The Fourier transform of the Moyal bracket}
Given two functions, $f(x)$ and $g(x)$, let $\hat f := \mathcal{F} f$ and $\hat g := \mathcal{F} g$ be their Fourier transforms and define 
\begin{equation}  \label{Ap80}
[\hat f, \hat g] := \mathcal{F}[f,g]\,,
\end{equation}
that is,
\begin{eqnarray}
 [\hat f, \hat g]_{(p)} &=& \frac1{(2\pi)^2}\, \int_{\RR^4} \D x\,e^{-i p x}\int_{\RR^8} \D u\, \D v\,f(x+u)\,g(x+v)\,S(u,v) \nonumber \\[2ex]
 & = & \frac1{(2\pi)^6}\,\int_{\RR^{20}} \D x\,\D u\,\D v\,\D q\,\D p\,e^{i(k+q- p) x + i q u + i k v}\,\hat f(q)\,\hat g(k)\,S(u,v) \nonumber  \\[2ex]  \label{Ap99}
  & = & \frac1{(2\pi)^2}\,\int_{\RR^8} \D q\,\D p\,\delta(k+q- p)\,\hat f(q)\,\hat g(k)\,\Sigma(q,k) \,,
\end{eqnarray}
with
$$ 	\Sigma(q,k) := \int_{\RR^8} \D u\,\D v\,e^{i q u + i k v}\,S(u,v) = - 2 i \,\sin\left(\frac{q\theta k}2 \right) \,, $$
where the expression (\ref{Ap7}) for $S(u,v)$ has been included. Then, replacing this into (\ref{Ap99}), we easily arrive at
\begin{equation}  \label{Ap98}
 [\hat f, \hat g]_{(k)} = \frac{i}{2\pi^2}\,\int_{\RR^8} \D q\,\D p\,\delta(p+q-k)\, \hat f(q)\,\hat g(p)\, \sin\left(\frac{q\theta p}2 \right) \,.
\end{equation}

\subsection*{A list of useful properties}
It is worth listing some properties that will be helpful for the developments throughout the paper, and whose proofs easily follow from the definitions:

\begin{description}
\item[1.-] The Moyal product is associative,
      $\quad  (f\ast g)\ast h = f\ast (g\ast h ) \,$,
\item[2.-] and fulfills the Jacobi identity
\begin{equation}  \label{Ap9}
 [[f, g], h ] +  [[g, h], f ] +  [[h, f], g ] = 0\,.
\end{equation}
\item[3.-] The equality concerning the overall integral
$\,\displaystyle{\int_{\RR^4} \D x\, \left(f \ast g\right)_{(x)} = \int_{\RR^4} \D x\, f(x) \,g(x)}\,$
is useful to prove three helpful equalities:
\begin{equation}  \label{Ap10a}
\int_{\RR^4} \D x\,f(x)\,\left[g,h\right]_{(x)} = \int_{\RR^4} \D x\,h(x)\,\left[f,g\right]_{(x)}  \,,\qquad \int_{\RR^4}\D x\,\left[f,g\right]_{(x)} = 0\,.  
\end{equation} 
and
\begin{equation}  \label{Ap10b}
\int_{\RR^4} \D x\, f(x)\,\left[g, \delta_z \right]_{(x)}  =\left[f, g \right]_{(z)}\,,\qquad {\rm where} \qquad \delta_y(x):=\delta(x-y)\,.
\end{equation}   
\item[4.-] The Leibniz rule for the derivative easily follows from (\ref{Ap7a}-\ref{Ap7}): 
\begin{eqnarray}  \label{Ap10c}
\partial_a \left(g\ast f \right) &=& \left(\partial_a g \right)\ast f + g \ast \left(\partial_a f \right)\,,   \\[1.5ex] \label{Ap10ca}
\partial_a \left[g, f \right] &=&  \left[\partial_a g, f \right] + \left[g, \partial_a f \right]\,,   \\[1.5ex] \label{Ap10d}
D_a \left[g, f \right]  &=& \left[D_a g, f \right]+ \left[g,D_a f\right]\,.
\end{eqnarray}  
\item[5.-] Let $x$ and $z$ be independent variables, then
\begin{equation}  \label{Ap10e}
  \int \D x\,f(x)\,\left(D_a\delta_z\right)_{(x)} = - \left(D_a f\right)_{(z)}\,.
\end{equation}
\item[6.-] Let $\;\displaystyle{ \xi =\frac{s-1}s\,y +\frac{x}s }\,$, then 
\begin{equation}  \label{Ap10f}
  \int \D y\,f(y)\, \left(D_a\delta_y\right)_{(\xi)} = s^{n+1}\,f_{|a}(x) - i\,\int\D u\,\D v \tilde{M}\left(u,\frac{v}s\right)\, A_a(x+u)\,f(x+v)\,.
\end{equation}
\item[7.-] 
\begin{equation}  \label{Ap10g}
  [f,1]_{(y)} = 0 \qquad {\rm and} \qquad 
	\left[f, \delta_x\right]_{(y)} = - \left[f, \delta_y\right]_{(x)}\,.
\end{equation}
\item[8.-] Let $\;\xi= \alpha y + \beta x\,$, then
\begin{equation}  \label{Ap10h}
 \frac{\partial\;}{\partial y^a}\,\left[f, \delta_y\right]_{(\xi)} = \alpha\,\left[f_{|a}, \delta_y\right]_{(\xi)} + (\alpha -1)\,\left[f, \delta_{y|a}\right]_{(\xi)}\,.
\end{equation}
\item[9.-]
\begin{equation}  \label{Ap10i}
\left(D_a\delta_y\right)_{(x)} = - \left(D_a\delta_x\right)_{(y)} = \delta_{|a}(x-y)- i\,[A_a,\delta_y]_{(x)}\,.
\end{equation} 
\item[10.-] 
\begin{equation}  \label{Ap10j}
 [D_a, D_b] f = - i\,[F_{ab},f]\,.
\end{equation}
\item[11.-] 
\begin{equation}  \label{Ap10k}
\int\D y\,f(y)\,\left(D_a g\right)_{(y)} = - \int\D y\,g(y)\,\left(D_a f\right)_{(y)} + \int\D y\,\partial_a \left(f(y) g(y) \right) \,,
\end{equation}
and, provided that $f(y) g(y) \rightarrow 0$ at infinity, the last term on the right hand side vanishes.
\end{description}

\section*{Appendix B: The symplectic form and the Hamiltonian up to $O(2)$}
The perturbative expansion of (\ref{M3}) demands the previous expansion of (\ref{NC1}) and (\ref{e108}). Keeping the two lowest orders only, we have that
\begin{eqnarray*}
\hat F^{ab} &=& 2 i \, k^{[a} \hat Z^{b]} + i \left(-[\hat Z^a, \hat Z^b] + \frac{2 k^{[a} \hat\Phi^{b]}}{k^c k_c} \right) + O(2)\,,  \\[2ex]
\hat\Phi^c & = & 2 k^a \,[\hat Z_a,\hat Z^c] - [\hat Z_a, k^c \hat Z^a] + O(2) \,.
\end{eqnarray*}
Combining them and using the FT of (\ref{Ap9}), one easily arrives at
\begin{equation} \label{m9z}
\hat F^{ab} = 2 i \, k^{[a} \hat Z^{b]} + i \, [\hat Z^c, \hat Z^d] \,
\left( - \delta_c^{[a}  +\frac4{k^e k_e}\, k_c k^{[a}\right)\, \delta_d^{b]} - \frac{2 i}{k^e k_e}\,
 \left[k^{a}\hat Z_c, k^{b} \hat Z^c \right] + O(2) \,.
\end{equation}

Straightforward computation that includes (\ref{e108z}) yields
$$\left[\hat Z^c, \hat Z^d \right]_{(k)} = \frac{i}{2 \pi^2}\,\sum_{\mu,\nu}\int_{\RR^6} \D \mq\,\D\Mp\,\delta(\mk - \mq-\Mp)\,W_\nu^c(\mq)\,W_\mu^d(\Mp)\,\delta(k^4-\nu|\mq| - \mu|\Mp|)\,\sin\left(\frac{q_\nu \theta p_\mu}2\right) $$
and
$$\left[k^a\hat Z_c, k^b \hat Z^c \right]_{(k)} = \frac{i}{2 \pi^2}\,\sum_{\mu,\nu}\int_{\RR^6} \D \mq\,\D\Mp\,\delta(\mk - \mq-\Mp)\,q^{[a}_\nu\,p^{b]}_\mu\,W_{\nu c}(\mq)\,W_\mu^c(\Mp)\,\delta(k^4-\nu|\mq| - \mu|\Mp|)\,\sin\left(\frac{q_\nu \theta p_\mu}2\right) \, ,$$
which, introduced in (\ref{m9z}), yields
\begin{eqnarray} 
\hat F^{ab} &=& 2 i \, k^{[a} \hat Z^{b]} + \frac{1}{\pi^2}\,\sum_{\mu,\nu}\int_{\RR^6} \D \mq\,\D\Mp\,W_{\nu c}(\mq)\,W_{\mu d}(\Mp)\,\delta(k - q_\nu-p_\mu) \nonumber\\[2ex] 
 & & \times \sin\left(\frac{q_\nu \theta p_\mu}2\right) \,\left\{\frac12 \,\eta^{d[b}\, \eta^{c a]} - \frac2{k^e k_e}\, k^{c} \eta^{d[b}\, k^{a]}  + \frac1{k_e k^e}\,\eta^{cd} \,q_\nu^{[a} p_\mu^{b]} \right\}+ O(2) \,.\label{m9y}
\end{eqnarray}
Using this, the second and third terms in equation (\ref{M3}) respectively become
\begin{eqnarray*}
-\lefteqn{\frac1{2\pi}\,\left[\int_\RR \frac{\D\tau}{\tau + i 0}\,T_{-\tau} \hat F^{ac},\hat A_c\right] = -\frac{i}{\pi}\,\sum_{\mu=\pm} \left[\frac{k_\mu^{[a} W_\mu^{c]}(\mk)}{k^4 -\mu |\mk|+i 0} , \hat{Z}_c\right]_{(k)} + O(2) } \\[2ex]
 &=& \frac{1}{2\pi^3}\,\sum_{\mu,\nu}\int_{\RR^6} \D \mq\,\D\Mp\,W_{\nu c}(\mq)\,W_{\mu b}(\Mp)\,\delta(\mk - \mq- \Mp) \,\sin\left(\frac{k \theta p_\mu}2\right) \,\frac{q_\nu^{[a} \eta^{b]c}}{k^4-\nu|\mq|-\mu|\Mp| + i 0} +O(2)
\end{eqnarray*}
and
\begin{eqnarray*}
\frac1{2\pi}\,\int_\RR \frac{\D\tau}{\tau + i 0}\,T_{-\tau} \left[\hat F^{ac},\hat A_c\right] &=& -\frac1{2\pi^3}\,\sum_{\nu,\mu} 
\int_{\RR^6} \D \mq\,\D\Mp\,W_{\nu c}(\mq)\,W_{\mu b}(\Mp)\,\sin\left(\frac{q_\nu \theta p_\mu}2\right) \, \\[2ex]
 & & \hspace*{-1em} \,\times \delta(\mk - \mq- \Mp) \,\frac{q_\nu^{[a} \eta^{b]c}}{k^4-\nu|\mq|-\mu|\Mp| + i 0} +O(2)\,,
\end{eqnarray*}
whereas the first term yields
\begin{eqnarray*} 
\frac1{2\pi} \int_\RR \D k^4 \,\hat F^{a4}(k) &=& \frac{i}{\pi}\,\sum_{\rho=\pm} k_\rho^{[a} W_\rho^{4]}(\mk) + \frac{1}{2\pi^3}\,\sum_{\mu,\nu}\int_{\RR^6} \D \mq\,\D\Mp\,W_{\nu c}(\mq)\,W_{\mu b}(\Mp)\,\delta(\mk - \mq- \Mp) \nonumber \\[2ex]  
 & & \hspace*{-4em}\times\sin\left(\frac{q_\nu \theta p_\mu}2\right) \,\left\{ \eta^{b[4}\,\left(\frac12 \, \eta^{a]c}  -\frac2{f^{(2)}_{\nu\mu}} \, ( q_\nu^{a]} + p_\mu^{a]}) \, ( q_\nu^c + p_\mu^c) \right) + \frac1{f^{(2)}_{\nu\mu}}\,\eta^{cb}\,q_\nu^{[a} p_\mu^{4]} \right\} + O(2)\,,
\end{eqnarray*}
where $\,f^{(2)}_{\nu\mu}:= |\mk|^2 - \left(\nu|\mq| + \mu |\Mp|\right)^2\,$. Putting them together and keeping only the symmetric part with respect to the exchange $\, (\nu, q^a,c) \leftrightarrow (\mu, p^a,b) \,$, equation (\ref{M3}) becomes
\begin{eqnarray} 
\hat P^a(k) &=& \frac{i}{\pi}\,\sum_{\rho=\pm} k_\rho^{[a} W_\rho^{4]}(\mk) + \frac1{4\pi^3}\,\sum_{\mu,\nu}\int_{\RR^6} \D \mq\,\D\Mp\,\delta(\mk - \mq-\Mp) \,W_{\nu c}(\mq)\,W_{\mu b}(\Mp) \times \nonumber \\[2ex]   \label{M4}
 & & \,\hspace*{-6em}\left\{ \sin\left(\frac{q_\nu \theta p_\mu}2\right) \,\left(m^{acb}_{ \nu\mu} - n^{acb}_{\;\,\nu\mu} + n^{abc}_{\;\,\mu\nu} \right) + \sin\left(\frac{k \theta p_\mu}2\right)\,n^{acb}_{\nu\mu} + \sin\left(\frac{k\theta q_\nu}2\right)\,n^{abc}_{\mu\nu}   \right\} + O(2) \,,
\end{eqnarray}
where $\,m^{acb}_{ \nu\mu}\,$ and  $\,n^{acb}_{ \nu\mu}\,$ stand for 
\begin{eqnarray} \label{M4a}
m^{acb}_{ \nu\mu}(\mk,\mq,\Mp) &=& \eta^{b[4} \eta^{a]c} + \frac2{f^{(2)}_{ \nu\mu}}\,\left[ \eta^{cb} \,q^{[a}_\nu p^{4]}_\mu - 2 (q_\nu^{[c} + p_\mu^{[c})\,\eta^{b][4}\,(q_\nu^{a]} + p_\mu^{a]})\right] \\[2ex]   \label{M4b}
{\rm and} \quad n^{acb}_{\nu\mu}(\mk,\mq,\Mp) & = &   \frac{1}{g_{\nu\mu}(k)+ i 0}\, q_\nu^{[a}\, \eta^{b]c}  \,,
\end{eqnarray}
with $\,g_{\nu\mu}(k):= k^4-\nu|\mq| -\mu|\Mp|\,$.

\subsection*{The symplectic form}
The first order of the symplectic form (\ref{Sp1b}) is
\begin{eqnarray}
 \omega_1 &=& \int_{\RR^4}\D k \,\left(\delta \hat P_1^a(-k)\wedge \delta\hat Z_a(k)+ \delta \hat P_0^a(k)\wedge \delta\hat A_{1 a}(-k) \right) \nonumber \\[2ex]  \label{Sp1bz}
  & = & \sum_\rho \int_{\RR^3} \D\mk\,\left(\delta \hat P_1^a(-k_\rho) -\frac{i}\pi\,k_\rho^{[d} \eta^{4]a} \, \int_\RR \D k^4\, \delta \hat A_{1 d}(-k) \right) \wedge \delta W_{\rho a}(\mk)\,.
\end{eqnarray}
From (\ref{M4}), it follows that
\begin{eqnarray}
\delta \hat P_1^a(-k_\rho) &=& \frac1{2 \pi^3}\,\sum_{\mu,\nu}\int_{\RR^6} \D \mq\,\D\Mp\,\delta(\mk + \mq+\Mp) \,W_{\mu b}(\Mp)\, \delta W_{\nu c}(\mq) \,\left\{ \sin\left(\frac{q_\nu \theta p_\mu}2\right) \, \right. \nonumber \\[2ex]   \label{M5}
 & & \left. \hspace*{-1em} \times  \left(m^{acb}_{ \nu\mu}- \tilde n^{acb}_{\rho\nu\mu} + \tilde n^{abc}_{\rho\nu\mu}\right) - \sin\left(\frac{k_\rho\theta p_\mu }2\right) \,\tilde n^{acb}_{\rho\nu\mu} - \sin\left(\frac{k_\rho \theta q_\nu}2\right) \,\tilde n^{abc}_{\rho\mu\nu} \right\}    \,,
\end{eqnarray}
where $\,\tilde n^{acb}_{\rho\nu\mu}\,$ stands for
$$ \tilde n^{acb}_{\rho\nu\mu}(\mk,\mq,\Mp) :=  - \frac1{f_{\rho\nu\mu} - i0}\,q_\nu^{[a}\eta^{b]c}   \,,$$
where $f_{\rho\nu\mu} = - \,g_{\nu\mu}(-k_\rho) = \rho |\mathbf{k}| + \nu|\mathbf{q}| + \mu |\mathbf{p}|\,.$ 

On the other hand, from (\ref{m11}), we have that 
\begin{equation} \label{M5a}
 \int_\RR \D k^4\,\delta \hat A_{1 d}(-k) =  \frac{i}{\pi^2}\,\sum_{\mu,\nu}\int_{\RR^6} \D \mq\,\D\Mp\,\delta(\mk + \mq+\Mp) \,W_{\mu b}(\Mp)\, \delta W_{\nu c}(\mq) \, \sin\left(\frac{q_\nu \theta p_\mu}2\right) \,h^{ecb}_{ \nu\mu} \eta_{ed}  \,,
\end{equation}
where $\,h^{ecb}_{ \nu\mu}\,$ is given by (\ref{m11a}).
Combining both equations (\ref{M5}-\ref{M5a}) with (\ref{Sp1bz}), we arrive at
\begin{eqnarray}
 \omega_1 &=&  \frac1{4 \pi^3}\,\sum_{\rho,\mu,\nu}\int_{\RR^9} \D \mk\,\D \mq\,\D\Mp\,\delta(\mk + \mq+\Mp) \,W_{\mu b}(\Mp)\, \delta W_{\nu c}(\mq) \wedge \delta W_{\rho a}(\mk) \,\nonumber \\[2ex] \label{M6}
  &  & \times\left\{ \sin\left(\frac{q_\nu \theta p_\mu}2\right) \,C^{acb}_{\rho\nu\mu}  - \sin\left(\frac{k_\rho\theta p_\mu }2\right) \,C^{cab}_{\nu\rho\mu} + \sin\left(\frac{k_\rho \theta q_\nu}2\right) \,B^{abc}_{\rho\mu\nu} \right\}
\end{eqnarray}
with
\begin{eqnarray}   \label{M6a}
 C^{acb}_{\rho\nu\mu} &=& m^{acb}_{ \nu\mu} + 2\,h^{ecb}_{\;\,\nu\mu} \eta_{ed} \,k_\rho^{[d} \eta^{4]a} + \tilde n^{cab}_{\nu\rho\mu} - \tilde n^{acb}_{\rho\nu\mu} + \tilde n^{abc}_{\rho \mu\nu} \\[2ex] \label{M6b}
 B^{abc}_{\rho\mu\nu} &=& - \tilde n^{abc}_{\rho\mu\nu} - \tilde n^{cba}_{\nu\mu\rho} = 0\,. 
\end{eqnarray}
In order to determine $\,C^{acb}_{\rho\nu\mu}\,$, we use that
$$ m^{acb}_{ \nu\mu} + 2\,h^{ecb}_{\;\,\nu\mu} \eta_{ed} \,k_\rho^{[d} \eta^{4]a} = \eta^{b[4} \eta^{a]c} + \frac1{f_{\rho\nu\mu}}\, \left(- q_\nu^a \eta^{cb} -  p_\mu^c \eta^{ab} +  q_\nu^b \eta^{ac} \right) - \frac12 \,\eta^{cb} \eta^{4a}\,,$$
where we have taken into account that $q^2_\nu=p^2_\mu= 0$ and $k^a_\rho + q^a_\nu + p^a_\mu = f_{\rho\nu\mu}\,\delta^a_4$. 
We also use that  
$$ \tilde n^{cab}_{\nu\rho\mu} - \tilde n^{acb}_{\rho\nu\mu} + \tilde n^{abc}_{\rho\mu\nu} = - \frac1{f_{\rho\nu\mu} - i0}\, \left(- q_\nu^a \eta^{cb} -  p_\mu^c \eta^{ab} +  q_\nu^b \eta^{ac} \right) + \frac12 \,\eta^{cb} \eta^{4a} - \eta^{b[4} \eta^{a]c}\,.$$
Therefore, $\,C^{acb}_{\rho\nu\mu} = 0 \,$, whence it follows that
$$\omega = \omega_0 + O(2)\,. $$

\subsection*{The Hamiltonian}
The first order of the Hamiltonian (\ref{S54_h_D}) can be decomposed into $$h_1 = h_{10}+ h_{11}\,,$$ with 
\begin{eqnarray}    \label{AH_h1}
     h_{10} &:=& i \int_{\mathbb{R}^4} \mathrm{d}k\,k^4\,\left\{\hat P^a_0(k)\hat Z_a(-k) + \hat P^a_0(k) \hat A_{1a}(-k) + \hat P^a_1(k) \hat Z_a(-k)\right\}\nonumber\\[1ex] \label{AH_h}
    h_{11} &:=&  \frac{1}{4\pi} \int_{\mathbb{R}^5} \mathrm{d}k \, \mathrm{d}\sigma \,\hat F^{ab}_0(k) \hat F_{1ab}(-\tilde k)  \,, 
\end{eqnarray}
where $\tilde k = (\mathbf{k},\sigma)$ and 
\begin{eqnarray}
    \hat F^{ab}_0(k) & = & 2i k^{[a}\hat Z^{b]}(k)\\
    \hat F^{ab}_1(k) & = & \frac{1}{\pi^2}\,\sum_{\mu,\nu}\int_{\RR^6} \D \mq\,\D\Mp\,W_{\nu c}(\mq)\,W_{\mu d}(\Mp)\,\delta(k - q_\nu-p_\mu) \nonumber\\
 & & \times \sin\left(\frac{q_\nu \theta p_\mu}2\right) \,\left\{\frac12 \,\eta^{d[b}\, \eta^{c a]} - \frac2{k^e k_e}\, k^{c} \eta^{d[b}\, k^{a]}  + \frac1{k_e k^e}\,\eta^{cd} \,q_\nu^{[a} p_\mu^{b]} \right\} \,.\label{AH_F1NS}
\end{eqnarray}
Keeping only the symmetric terms with respect to the change $(\nu,\mathbf{q},c) \leftrightarrow (\mu,\mathbf{p},d)$ and using the Dirac delta function, the right hand side of the latter becomes
\begin{eqnarray*}
    \hat F^{ab}_1(k) & = & \frac{1}{2\pi^2}\,\sum_{\mu,\nu}\int_{\RR^6} \D \mq\,\D\Mp\,W_{\nu c}(\mq)\,W_{\mu d}(\Mp)\,\delta(k - q_\nu-p_\mu) \nonumber\\
 & & \times \sin\left(\frac{q_\nu \theta p_\mu}2\right) \,\left\{\eta^{d[b}\, \eta^{a]c} + \frac2{f^{(2)}_{\nu\mu}}\left[\eta^{cd}q^{[a}_\nu p^{b]}_\mu - 2(q_\nu^{[c} + p_\mu^{[c})\,\eta^{d][b}\, (q^{a]}_\nu + p_\mu^{a]}) \right]   \right\}. \label{AH_F1S}
\end{eqnarray*}
Therefore, equation (\ref{AH_h}) can be written as
\begin{equation*}
h_{11} = \frac{i}{4\pi^3} \sum_{\rho\nu\mu} \int_{\mathbb{R}^9}\mathrm{d}\mathbf{q}\, \mathrm{d}\mathbf{p}\,\mathrm{d}\mathbf{k}\,W_{\nu c}(\mathbf{q})\, W_{\mu b}(\mathbf{p})\,W_{\rho d}(\mathbf{k})\,\delta(\mathbf{k+q+p})\,M^{cbd}_{\nu\mu\rho}(\mathbf{q,p,k})\,,
\end{equation*}
with
\begin{equation*}
    M^{cbd}_{\nu\mu\rho}(\mathbf{q,p,k}):=\sin\left(\frac{q_\nu \theta p_\mu}{2}\right)  k_{\rho a} \left\{\eta^{b[d}\, \eta^{a]c} + \frac2{f^{(2)}_{ \nu\mu}}\left[\eta^{cb}q^{[a}_\nu p^{d]}_\mu - 2(q_\nu^{[c} + p_\mu^{[c})\,\eta^{b][d}\, (q^{a]}_\nu + p_\mu^{a]}) \right] \right\}\,, 
\end{equation*}
Using (\ref{e108z}) and the facts that $k^d W_d = 0$, $k^2_\rho =0 $ and $k^a_\rho + q^a_\nu + p^a_\mu = f_{\rho\nu\mu} \delta^a_4$, we can simplify it further:
\begin{equation*}
    M^{cbd}_{\nu\mu\rho}(\mathbf{q,p,k}):=\sin\left(\frac{q_\nu \theta p_\mu}{2}\right) \left\{\eta^{d[b}\,k^{c]}_\rho + \frac{2\,k_{\rho a}}{f^{(2)}_{\nu\mu}}\left[\eta^{cb} q^{[a}_\nu p^{d]}_\mu + f_{\rho\nu\mu} \eta^{4[a}\left( p^c_\mu \eta^{d]b} - q^b_\nu \eta^{d]c}\right)  \right] \right\}\,. 
\end{equation*}
Moreover, it is worth remarking that  
$$M^{cbd}_{\nu\mu\rho}(\mathbf{q,p,k}) = M^{bcd}_{\mu\nu\rho}(\mathbf{p,q,k})\,.$$ 
Therefore, total symmetrization with respect to the indices $(\nu,\mathbf{q},c) \leftrightarrow (\mu,\mathbf{p},b) \leftrightarrow (\rho,\mathbf{k}, d)$, we finally obtain that
\begin{eqnarray}
    h_{11} &=& \frac{i}{12\pi^3} \sum_{\rho\nu\mu} \int_{\mathbb{R}^9}\mathrm{d}\mathbf{q}\, \mathrm{d}\mathbf{p}\,\mathrm{d}\mathbf{k}\,W_{\nu c}(\mathbf{q})\, W_{\mu b}(\mathbf{p})\,W_{\rho d}(\mathbf{k})\,\delta(\mathbf{k+q+p})\nonumber\\
    & &\qquad \qquad \qquad \times\left\{M^{cbd}_{\nu\mu\rho}(\mathbf{q,p,k}) + M^{cdb}_{\nu\rho\mu}(\mathbf{q,k,q}) + M^{bdc}_{\mu\rho\nu}(\mathbf{p,k,q})\right\}\,.
\end{eqnarray}

As for the term $h_{10}$, we have that 
\begin{eqnarray}
h_{10} & = &  -\frac{i}{4\pi^3}\sum_{\rho \nu \mu} \int_{\mathbb{R}^9} \mathrm{d}\mathbf{q}\,\mathrm{d}\mathbf{p}\,\mathrm{d}\mathbf{k} \,\delta(\mathbf{k + q + p}) W_{\nu c}(\mathbf{q})\, W_{\mu b}(\mathbf{p})\, W_{\rho d}(\mathbf{k})\times \nonumber \\[1ex]
 & & \left\{\sin\left(\frac{q_\nu \theta p_\mu}{2}\right)\left[ \rho|\mathbf{k}|  \left( m^{dcb}_{\rho\nu\mu} - \tilde n^{dcb}_{\rho\nu\mu} + \tilde n^{dbc}_{\rho\mu\nu} \right) - 2\left(\nu|\mathbf{q}| + \mu|\mathbf{p}|\right) k^{[a}_\rho \eta^{4] d} h^{ecb}_{\rho\nu\mu} \eta_{e a}\right] -  \right.  \nonumber \\[1ex]
 & & \left.  \sin\left(\frac{k_\rho \theta p_\mu}{2}\right) \rho |\mathbf{k}| \tilde n^{dcb}_{\rho\nu\mu} - \sin\left(\frac{k_\rho \theta q_\nu}{2}\right) \rho |\mathbf{k}| \tilde n^{dbc}_{\rho \mu \nu} \right\} \nonumber
\end{eqnarray}
that, by symmetrising with respect to the indexes $(\nu,\mathbf{q},c) \leftrightarrow (\mu,\mathbf{p},b) \leftrightarrow (\rho,\mathbf{k}, d)$, we get
\begin{eqnarray}
    h_{10} &=& - \frac{i}{12\pi^3} \sum_{\rho\nu\mu} \int_{\mathbb{R}^9}\mathrm{d}\mathbf{q}\, \mathrm{d}\mathbf{p}\,\mathrm{d}\mathbf{k}\,W_{\nu c}(\mathbf{q})\, W_{\mu b}(\mathbf{p})\,W_{\rho d}(\mathbf{k})\,\delta(\mathbf{k+q+p})\nonumber\\
    & &\qquad \qquad \qquad \times\left\{\tilde{M}^{cbd}_{\nu\mu\rho}(\mathbf{q,p,k}) + \tilde{M}^{cdb}_{\nu\rho\mu}(\mathbf{q,k,q}) + \tilde{M}^{bdc}_{\mu\rho\nu}(\mathbf{p,k,q})\right\}\,,
\end{eqnarray}
where
\begin{eqnarray*}
\tilde{M}^{cbd}_{\nu\mu\rho}(\mathbf{q,p,k}):=\sin\left(\frac{q_\nu\theta p_\mu}{2}\right) \left\{\rho|\mathbf{k}|\left(m^{dcb}_{\rho\nu\mu} - \tilde n^{dcb}_{\rho\nu\mu} - \tilde n^{dbc}_{\rho \mu \nu}\right) -\left(\nu|\mathbf{q} + |\mu| \mathbf{p}\right) \left[\tilde n^{cdb}_{\nu\rho\mu} + 2 k_\rho^{[a} \eta^{4]d} h^{ecb}_{\rho \nu \mu} \eta_{ea} \right]\right\}\,.
\end{eqnarray*}
Now, if we use the relation $C^{acb}_{\rho \nu \mu} = 0$, we can simplify $\tilde{M}^{cbd}_{\nu\mu\rho}$ so that 
\begin{equation*}
\tilde{M}^{cbd}_{\nu\mu\rho}(\mathbf{q,p,k}) = \sin\left(\frac{q_\nu\theta p_\mu}{2}\right)\left\{- f_{\rho\nu\mu} \tilde n^{cdb}_{\nu\rho\mu} - 2 f_{\rho\nu\mu} k^{[a}_\rho \eta^{4]d} h^{ecb}_{\rho \nu \mu} \eta_{ea}\right\}\,.
\end{equation*}
The first term is
\begin{equation*}
- f_{\rho\nu\mu} \tilde n^{cdb}_{\nu\rho\mu} = \eta^{d[b}k^{c]}_\rho\,,
\end{equation*}
and the second term is 
\begin{equation*}
- 2 f_{\rho\nu\mu} k^{[a}_\rho \eta^{4]d} h^{ecb}_{ \nu \mu} \eta_{ea} = \frac{2\,k_{a\rho}}{f^{(2)}_{ \nu\mu}} \eta^{4[a}\left(p^c_\mu \eta^{d] b} - q^b_\nu \eta^{d]c}\right) - \frac{k_{\rho a} \eta^{cb}}{f^{(2)}_{ \nu\mu}} f_{\rho\nu\mu} \eta^{4[a} \left(p^{d]}_\mu - q^{d]}_\nu\right)\,.
\end{equation*}
To further simplify the right-hand side, we will use (\ref{e114v}) together with $q^a_\nu + p^a_\mu + k^a_\rho = - f_{\rho\nu\mu} \eta^{4a}$and $k^2_\rho = 0$. Then, we finally arrive at 
\begin{equation*}
- 2 f_{\rho\nu\mu} k^{[a}_\rho \eta^{4]d} h^{ecb}_{ \nu \mu} \eta_{ea} = \frac{2\,k_{\rho a}}{f^{(2)}_{ \nu\mu}}\left[\eta^{cb} q^{[a}_\nu p^{d]}_\mu + f_{\rho\nu\mu} \eta^{4[a}\left( p^c_\mu \eta^{d]b} - q^b_\nu \eta^{d]c}\right)\right]\,.
\end{equation*}
Thus we find that $\tilde M^{cbd}_{\nu \mu \rho}$ is nothing but $M^{cbd}_{\nu \mu \rho}$, rewritten in a different way. Then, combining $h_{10}$ and $h_{11}$, we get that  $h_1 = 0$; therefore,
\begin{equation*}
h = h_0 + O(2)\,.    
\end{equation*}

\subsection*{The potentials in terms of the canonical variables $\mU$ and $\mU^\dagger$}
We start substituting  $\,W_\nu^a(\mq) = \delta^a_j\,U^j(\mq)$ in the potentials (\ref{m11}). Then we exchange the dummy variables $\mq$ and $\Mp$ when necessary, and use that $\,p^b_-(-\Mp) = - p^b_+(\Mp)\,$ and that
$$  h^{acb}(-q_\nu,-p_\mu) = - h^{acb}(q_\nu,p_\mu)  \qquad \mbox{and} \qquad h^{acb}(q_\nu,p_\mu) = h^{abc}(p_\mu,q_\nu) \,,$$
together with equations (\ref{NC_canVar}) and (\ref{NC_canVar_2}) to obtain
\begin{eqnarray*}
\hat A^a(k) &=& \hat Z^a(k) +  \frac{i}{2\pi} \,\int_{\RR^6} \frac{\D\mq\, \D\Mp}{\sqrt{|\mq|\,|\Mp|}}\,\sin\left(\frac{q_+\theta p_+}{2}\right) \,\left\{ 2\,\delta(k^b-q^b_+ + p^b_+)\,h^{ajl}(q_+,-p_+)\,U_j(\mq)\,U_l^\dagger(\Mp) \right.
 +   \\[2ex]
 & & \left. h^{ajl}(q_+,p_+)\,\left[\delta(k^b-q^b_+- p^b_+)\, U_j(\mq)\,U_l(\Mp) - \delta(k^b+q^b_+ + p^b_+)\, U_j^\dagger(\mq)\,U_l^\dagger(\Mp) \right]  \right\}  +O(2)  \,,
\end{eqnarray*}
with
$$ \hat{\mathbf{Z}} = \sqrt{\frac{\pi}{|\mk|}}\,\left\{\mU(\mk)\,\delta(k^4-|\mk|) + \mU^\dagger(\mk)\,\delta(k^4+|\mk|) \right\}  
\,, \qquad \hat Z^4 = 0 $$
and $h^{acb}(q_\nu,p_\mu)$ given by (\ref{m11a}). 

Alternatively the potentials can also be written as 
\begin{eqnarray}  
\hat A^a(k) &=& \hat Z^a(k) +  \frac{i}{2\pi} \,\int_{\RR^6}\frac{\D\mq\, \D\Mp}{\sqrt{|\mq|\,|\Mp|}}\,\sin\left(\frac{q_+\theta p_+}{2}\right) \,\left\{ 2\,\delta(k^b-q^b_+ + p^b_+)\,S^a(\mq,\Mp)  + \right. \nonumber \\[2ex]  \label{PotsApp}
 & & \left. \delta(k^b-q^b_+- p^b_+)\,R^a(\mq,\Mp) - \delta(k^b+q^b_+ + p^b_+)\, R^{\dagger\,a}(\mq,\Mp)   \right\}  +O(2)  \,,
\end{eqnarray}
where
$$ R^a(\mq,\Mp):= \frac1{\mk^2-(|\mq|+|\Mp|)^2}\,\left[\Mp\cdot\mU(\mq)\,\delta^a_l\,U^l(\Mp) - \mq\cdot\mU(\Mp)\,\delta^a_l\,U^l(\mq) - \frac12\,\mU(\mq)\cdot\mU(\Mp)\,\left(p_+^a - q_+^a\right) \right] $$
and
$$ S^a(\mq,\Mp):= \frac{-1}{\mk^2-(|\mq|-|\Mp|)^2}\,\left[\Mp\cdot\mU(\mq)\,\delta^a_l\,U^{\dagger\,l}(\Mp) + \mq\cdot\mU^\dagger(\Mp)\,\delta^a_l\,U^l(\mq)- \frac12\,\mU(\mq)\cdot\mU^\dagger(\Mp)\,\left(p_+^a+q_+^a\right) \right] \,.$$

%% file: NC1-Main.bbl
\providecommand{\href}[2]{#2}\begingroup\raggedright\begin{thebibliography}{10}

\bibitem{Heredia2023_1}
C.~Heredia and J.~Llosa, ``Infinite derivatives vs integral operators. {T}he
  {M}oeller-{Z}wiebach paradox,''
  \href{http://dx.doi.org/10.48550/ARXIV.2302.00684}{{\em arXiv: 2302.00684
  [hep-th]} (2023) }.

\bibitem{Carlsson2016}
M.~Carlsson, H.~Prado, and E.~G. Reyes, ``Differential equations with
  infinitely many derivatives and the {B}orel transform,''
  \href{http://dx.doi.org/10.1007/s00023-015-0447-4}{{\em Annales Henri
  Poincar{\'e}} {\bfseries 17} no.~8, (2016) 2049--2074}.

\bibitem{Heredia2022}
C.~Heredia and J.~Llosa, ``Nonlocal {L}agrangian fields: {N}oether's theorem
  and {H}amiltonian formalism,''
  \href{http://dx.doi.org/10.1103/physrevd.105.126002}{{\em Physical Review D}
  {\bfseries 105} no.~12, (Jun, 2022) }.

\bibitem{Heredia_PhD}
C.~Heredia, ``Nonlocal {L}agrangian formalism,''
  \href{http://dx.doi.org/https://doi.org/10.48550/arXiv.2304.10562}{{\em
  arXiv: 2304.10562 [hep-th]} (2023) }.

\bibitem{Llosa1994}
J.~Llosa and J.~Vives, ``Hamiltonian formalism for nonlocal lagrangians,''
  \href{http://dx.doi.org/10.1063/1.530492}{{\em Journal of Mathematical
  Physics} {\bfseries 35} (06, 1994) 2856--2877}.

\bibitem{Gomis2001}
J.~Gomis, K.~Kamimura, and J.~Llosa, ``Hamiltonian formalism for space-time
  noncommutative theories,''
  \href{http://dx.doi.org/10.1103/physrevd.63.045003}{{\em Physical Review D}
  {\bfseries 63} no.~4, (Jan, 2001) }.

\bibitem{Heredia1}
C.~Heredia and J.~Llosa, ``Energy-momentum tensor for the electromagnetic field
  in a dispersive medium,''
  \href{http://dx.doi.org/10.1088/2399-6528/abfd14}{{\em Journal of Physics
  Communications} {\bfseries 5} no.~5, (May, 2021) 055003}.

\bibitem{Heredia2}
C.~Heredia and J.~Llosa, ``Non-local {L}agrangian mechanics: {N}oether's
  theorem and {H}amiltonian formalism,''
  \href{http://dx.doi.org/10.1088/1751-8121/ac265c}{{\em Journal of Physics A:
  Mathematical and Theoretical} {\bfseries 54} no.~42, (Sep, 2021) 425202}.

\bibitem{Tomboulis2015}
E.~Tomboulis, ``Nonlocal and quasilocal field theories,''
  \href{http://dx.doi.org/10.1103/physrevd.92.125037}{{\em Physical Review D}
  {\bfseries 92} no.~12, (Dec., 2015) }.

\bibitem{JaumeGomis2000}
J.~Gomis and T.~Mehen, ``Space-time noncommutative field theories and
  unitarity,''
  \href{http://dx.doi.org/https://doi.org/10.1016/S0550-3213(00)00525-3}{{\em
  Nuclear Physics B} {\bfseries 591} no.~1, (2000) 265--276}.

\bibitem{Gomis2001_2}
J.~Gomis, K.~Kamimura, and T.~Mateos, ``Gauge and {BRST} generators for
  space-time non-commutative {U}(1) theory,''
  \href{http://dx.doi.org/10.1088/1126-6708/2001/03/010}{{\em Journal of High
  Energy Physics} {\bfseries 2001} no.~03, (Mar, 2001) 010--010}.

\bibitem{Naser2015}
N.~Ahmadiniaz, O.~Corradini, D.~D'Ascanio, S.~Estrada-Jim\'enez, and P.~Pisani,
  ``{Noncommutative {U}(1) gauge theory from a worldline perspective},''
  \href{http://dx.doi.org/10.1007/JHEP11(2015)069}{{\em JHEP} {\bfseries 11}
  (2015) 069}.

\bibitem{Kupriyanov2020}
V.~G. Kupriyanov and P.~Vitale, ``A novel approach to non-commutative gauge
  theory,'' \href{http://dx.doi.org/10.1007/jhep08(2020)041}{{\em Journal of
  High Energy Physics} {\bfseries 2020} no.~8, (Aug, 2020) }.

\bibitem{Kupriyanov2021}
V.~G. Kupriyanov, M.~Kurkov, and P.~Vitale, ``$\kappa$-{M}inkowski-deformation
  of {U}(1) gauge theory,''
  \href{http://dx.doi.org/10.1007/jhep01(2021)102}{{\em Journal of High Energy
  Physics} {\bfseries 2021} no.~1, (Jan, 2021) }.

\bibitem{Douglas_NCFT}
M.~R. Douglas and N.~A. Nekrasov, ``Noncommutative field theory,''
  \href{http://dx.doi.org/10.1103/RevModPhys.73.977}{{\em Rev. Mod. Phys.}
  {\bfseries 73} (Nov, 2001) 977--1029}.

\bibitem{Kupriyanov2020_2}
V.~G. Kupriyanov, ``Non-commutative deformation of {C}hern{\textendash}{S}imons
  theory,'' \href{http://dx.doi.org/10.1140/epjc/s10052-019-7573-y}{{\em The
  European Physical Journal C} {\bfseries 80} no.~1, (Jan, 2020) }.

\bibitem{Marija2023}
M.~D. {\'C}iri{\'c}, D.~Dordevi{\'c}, D.~Go{\v c}anin, B.~Nikoli{\'c}, and
  V.~Radovanovi{\'c}, ``Noncommutative {SO}(2,3) gauge theory of gravity,''
  \href{http://dx.doi.org/10.1140/epjs/s11734-023-00833-5}{{\em The European
  Physical Journal Special Topics} (2023) }.

\bibitem{Trampetic2023}
J.~Trampeti{\'c} and J.~You, ``Revisiting {NCQED} and scattering amplitudes,''
  \href{http://dx.doi.org/10.1140/epjs/s11734-023-00837-1}{{\em The European
  Physical Journal Special Topics} (2023) }.

\bibitem{Seiberg1999}
N.~Seiberg and E.~Witten, ``String theory and noncommutative geometry,''
  \href{http://dx.doi.org/10.1088/1126-6708/1999/09/032}{{\em Journal of High
  Energy Physics} {\bfseries 1999} no.~09, (Sep, 1999) 032--032}.

\bibitem{Pons2011}
J.~M. Pons, ``Noether symmetries, energy{\textendash}momentum tensors, and
  conformal invariance in classical field theory,''
  \href{http://dx.doi.org/10.1063/1.3532941}{{\em Journal of Mathematical
  Physics} {\bfseries 52} no.~1, (Jan, 2011) 012904}.

\bibitem{BELINFANTE1940}
F.~Belinfante, ``On the current and the density of the electric charge, the
  energy, the linear momentum and the angular momentum of arbitrary fields,''
  \href{http://dx.doi.org/https://doi.org/10.1016/S0031-8914(40)90091-X}{{\em
  Physica} {\bfseries 7} no.~5, (1940) 449--474}.

\bibitem{rosenfeld1940}
L.~Rosenfeld, {\em Sur le tenseur d'impulsion-{\'e}nergie}.
\newblock Acad{\'e}mie royale de Belgique. Classe des sciences. M{\'e}moires.
  Collection in-8$\,^{\circ}$. Tome 18. Fasc. 6. Palais des acad{\'e}mies,
  1940.

\bibitem{Moeller2002}
N.~Moeller and B.~Zwiebach, ``Dynamics with infinitely many time derivatives
  and rolling tachyons,''
  \href{http://dx.doi.org/10.1088/1126-6708/2002/10/034}{{\em Journal of High
  Energy Physics} {\bfseries 2002} no.~10, (Oct, 2002) 034--034}.

\bibitem{Luca2022}
L.~Buoninfante, Y.~Miyashita, and M.~Yamaguchi, ``{Topological defects in
  nonlocal field theories},''
  \href{http://dx.doi.org/10.1007/JHEP11(2022)104}{{\em JHEP} {\bfseries 11}
  (2022) 104}.

\bibitem{Heredia2020v1}
C.~Heredia and J.~Llosa, ``Energy-momentum tensor for the electromagnetic field
  in a dispersive medium as an application of noether theorem,''
  \href{http://dx.doi.org/10.48550/ARXIV.2002.12725}{{\em arXiv: 2002.12725v1
  (physics.class-ph)} (2020) }.

\bibitem{Choquet1982}
Y.~Choquet-Bruhat, {\em Analysis, manifolds, and physics}.
\newblock North-Holland, Amsterdam, rev.~ed., 1982.

\bibitem{Vladimirov}
V.~Vladimirov, {\em Equations of mathematical physics}.
\newblock Mir, 1984.

\bibitem{Dirac1964}
P.~Dirac, {\em Lectures on quantum mechanics}.
\newblock Belfer Graduate School of Science. Monographs series. Belfer Graduate
  School of Science, Yeshiva University, 1964.

\bibitem{zeidler2008}
E.~Zeidler, {\em Quantum Field Theory II: Quantum Electrodynamics: A Bridge
  between Mathematicians and Physicists}.
\newblock Quantum Field Theory. Springer Berlin Heidelberg, 2008.

\bibitem{Biswas2012}
T.~Biswas, E.~Gerwick, T.~Koivisto, and A.~Mazumdar, ``Towards singularity- and
  ghost-free theories of gravity,''
  \href{http://dx.doi.org/10.1103/physrevlett.108.031101}{{\em Physical Review
  Letters} {\bfseries 108} no.~3, (Jan, 2012) }.

\bibitem{Calcagni2018}
G.~Calcagni, L.~Modesto, and G.~Nardelli, ``Initial conditions and degrees of
  freedom of non-local gravity,''
  \href{http://dx.doi.org/10.1007/JHEP05(2018)087}{{\em Journal of High Energy
  Physics} {\bfseries 2018} no.~5, (2018) 87}.

\bibitem{Kol2021}
I.~Kol{\'a}{\v r} and J.~Boos, ``Retarded field of a uniformly accelerated
  source in nonlocal scalar field theory,''
  \href{http://dx.doi.org/10.1103/physrevd.103.105004}{{\em Physical Review D}
  {\bfseries 103} no.~10, (May, 2021) }.

\bibitem{Capozziello2022_NS}
A.~Acunzo, F.~Bajardi, and S.~Capozziello, ``Non-local curvature gravity
  cosmology via {N}oether symmetries,''
  \href{http://dx.doi.org/https://doi.org/10.1016/j.physletb.2022.136907}{{\em
  Physics Letters B} {\bfseries 826} (2022) 136907}.

\end{thebibliography}\endgroup
